\documentclass[conference,12pt, onecolumn,a4paper]{IEEEtran}
\makeatletter
\def\ps@headings{%
\def\@oddhead{\mbox{}\scriptsize\rightmark \hfil \thepage}%
\def\@evenhead{\scriptsize\thepage \hfil \leftmark\mbox{}}%
\def\@oddfoot{}%
\def\@evenfoot{}}
\makeatother
\usepackage{graphics,color} 
\usepackage{subfigure}
\usepackage{algorithmic}
\usepackage{algorithm,amssymb,url}
\usepackage{epsfig}

\usepackage[cmex10]{amsmath}

\newtheorem{corollary}{Corollary}
\newtheorem{proposition}{Proposition}
\def\Prf{\vspace{2ex}\noindent{\bf Proof: }}

\def\endpf{\hfill{$\blacksquare$}}
\newcommand {\beq} {\begin{equation}}

\newcommand{\fdp}[1]{\textcolor{black}{#1}}
\newcommand{\f}[1]{\textcolor{black}{#1}}

\newcommand {\eeq} {\end{equation}}
\newcommand {\barr} {\begin{array}}
\newcommand {\earr} {\end{array}}
\newcommand {\bear} {\begin{eqnarray}}
\newcommand {\eear} {\end{eqnarray}}
\newcommand {\bears} {\begin{eqnarray*}}
\newcommand {\eears} {\end{eqnarray*}}
\newtheorem{definition}{Definition}

\newtheorem{thm}{Theorem}
\newtheorem{remark}{Remark}

\begin{document}

\IEEEoverridecommandlockouts

\title{\f{Incentive Mechanisms based on Minority Game in Heterogeneous DTNs}}
\author{Wissam Chahin$^\star$\thanks{$^\star$CERI/LIA, University of Avignon, 339, chemin des Meinajaries,Avignon, France.  $^\dagger$ University of California at Berkeley, California-94720. $^{\ddagger}$CREATE-NET, via Alla Cascata 56 c, 38100 Trento, Italy.},
Habib B.A. Sidi$^\star$, Rachid El-Azouzi$^\star$$^\dagger$
Francesco De Pellegrini$^{\ddagger}$
Jean Walrand$^\dagger$
}

\maketitle

\begin{abstract}
In this paper we design an incentive mechanism for heterogeneous Delay Tolerant Networks (DTNs).
The proposed mechanism tackles a core problem of such systems: how to induce coordination
of DTN relays in order to achieve a target performance figure, e.g., delivery probability or end-to-end delay,
under a given constraint in term of network resources, e.g., number of active nodes or energy
consumption. Also, we  account for the realistic case when the cost for taking part in the forwarding
process varies with the devices' technology or the users' habits. Finally, the scheme is truly
 applicable to DTNs since it works with no need for end-to-end connectivity.

In this context, we first introduce the basic coordination mechanism leveraging the notion
of a Minority Game. In this game, relays compete to be in the population minority and
their utility is defined in combination with a rewarding mechanism. The rewards in turn configure as
a control by which the network operator controls the desired operating point for the DTN. To this aim,
we provide a full characterization of the equilibria of the game in the case of heterogeneous DTNs.
Finally, a learning algorithm based on stochastic approximations provably drives the system to the equilibrium
solution without requiring perfect state information at relay nodes or at the source node and without using
end-to-end communications to implement the rewarding scheme. We provide extensive numerical
results to validate the proposed scheme.

\end{abstract}

\section{Introduction}

Delay Tolerant Networks (DTNs) are designed to sustain communications with no need for persistent connectivity. As such, they configure as a mean to exchange
data without the help of dedicated network infrastructure. As described in related literature~\cite{ElAzouzi2012,giovani}, the way to overcome disconnections
in such systems is by message replication. The replication mechanism resembles an epidemics where the infection is represented by the data generated by a
source node and the infected nodes are those which fetch data from peer nodes.

In practice, DTNs composed by mobile devices could include smartphones, tablets or other devices using multiple wireless interfaces. For such devices, OS APIs are available to program dedicated applications for direct data exchange with peer nodes during radio range contacts.

Using such a flexible mechanism for communication, DTN networks can use mobile terminals as information carriers: as such, they can be employed to off-load infrastructure networks during critical demand situations. This is the case of modern flashcrowds or similar events: they represent a peculiar novel type of hotspot scenario were infrastructured networks are overused. In these events, due to the high densities of user-generated contents related to a specific geographic location, the operator faces massive demand on the limited spectrum of infrastructure networks, therefore leading to deteriorating wireless quality for all subscribers. 
The solution that is classically adopted by operators in this context is to deploy contingency sites to increase coverage, with the incurred costs for installation, removal and overall network engineering for interference mitigation. 

These contingency solutions are designed for specific events: in that scenario, the appeal of DTNs is precisely  that they are flexible in that users act directly as relays thus contributing to reducing infrastructure costs while ensuring good coverage. Furthermore, users can keep data exchange local, e.g., event-related information can be disseminate immediately among the participants with no need to outreach remote servers.  This architecture is thus expected to reduce network costs, while offering a new service that may interest a large community of users.

In order to make a DTN for local data exchange truly operational, several technical issues need to be tackled in order. To date, the most attention from the research community has been attracted on how to route messages reliably towards the intended destination(s). Replication of the original message by the so called epidemic routing protocol ensures that at least some copy will reach the destination node with high probability minimizing the delay to reach the intended destinations. In turn, the standard optimization problem becomes how to  maximize the delivery probability under constraints on the resources spent to forward it to the destination. Several papers have further extended the possible optimization to the use of activation and/or forwarding control at relays \cite{neglia06chants}.

In this paper we rather focus on a fundamental aspect that is usually overlooked in DTN literature, namely, the {\em autonomous activation} problem. DTN performance optimization is usually made under the implicit assumption that relays are willing to cooperate with the source node. But, core point is whether owners of relay devices, e.g., either smartphones or tables, are willing to have battery depleted to sustain DTNs communications. In turn, massive de-activation of relays becomes a core threat which hinders any possible attempt to optimize network performance.

To this respect, a fundamental observation is that relays may have heterogeneous energy requirements and so does the cost for activation they experience \cite{6098189,wiopt10}. In fact, there are two main sources of heterogeneity that we need to model. First, DTN nodes may belong to different categories, e.g., PDA, laptop, mobile and/or have related communication/energy-autonomy features. Those are determined by transmission range, mobility, memory, energy capacity and active radio interface such as WiFi and Bluetooth. Second, heterogeneity also includes differences in behavior of devices' owners\cite{dtn-heterog}. This covers a realistic scenario in which the cost experienced by users to activate is related to their activities and the priority they give to  tasks other than relay operations, e.g., facebook updates or phone calls may be prioritized over relaying activities depending on the user profile.

{\noindent \em Contribution of the paper: } The main contribution proposed lies in the design of a credit-based mechanism for relay participation in DTNs. Our mechanism design attains a twofold objective. First, the decision to participate to relaying or not is taken autonomously by relays according to the incentive scheme, i.e., the choice to activate or not depends on the local decision taken by the device and no centralized signal exchange is required. Second, since incentives engender a competition among relays that play strategies on their activation, they can be driven to attain a desired operating point for the DTN. Such an operating point, in turn, is precisely the solution of a joint optimization problem involving the number of active relays.

Thus, the reward offered is a {\em control variable} by which the source can tune the number of relays taking part to the forwarding process. To do so, we define a specific utility structure as linear combination of the success probability of a relay and the energy cost for activation. The success of a tagged relay depends on the number of active opponents met: the bigger the number of relays participating to the message delivery, the higher the delivery probability for the message, but indeed the less the chance for the tagged relay to receive a reward from the system. Furthermore, we will consider different information scenarios: message sources may have different information about the energy cost of relays. In the first scenario, the heterogeneity of energy is  based on the type of device (e.g.,  tablet, iPad, smartphone, laptop, access point). Sources can identify the type of  devices at each contact, thus the game  is  played under complete information on the energy cost of each class.   In the second scenario, we associate the energy cost to the activity of devices' owners or their profile.\footnote{In view of the reward mechanism proposed, this scenario can also reflect the importance for a tagged user to use the DTN to exchange its own messages since it can act as both a relay and a source.}

More in detail, our approach is grounded into the theory of the Minority Game (MG) \cite{Esteban04}  which tunes performance of competing relays and welfare of the DTN (number of message copies and delivery message). We thoroughly investigate the properties of our coordination game in which relays compete to be in the population minority.

\begin{remark} A peculiar requirement for our credit exchange mechanism is that it does not require end-to-end communications.  Due to the fact that feedback messages in DTNs may incur into large delays, in fact, the exchange of credits between relays should not require feedback messages. In order to overcome lack of feedback, our solution adopts a learning algorithm based on stochastic approximations. This algorithm performs in distributed fashion and operates based on local estimates performed by relays. As such, it is fully decentralized  with no need to reconstruct full state information neither at the source nor at relays. 
\end{remark}

\textbf{Background on Minority Game and incentive mechanisms:}\\
\f{{\em The Minority Game.} The standard minority game studies how individuals of a population of 
competing agents may reach a form of coordination  when sharing resources for which the utility decreases in the number of players. } 
The MG was first introduced in literature as a simplification of the El Farol Bar's attendance problem \cite{Esteban04,willemien07}. In the El Farol bar problem  \cite{gintis} $N$ users decide independently whether to go or not to the unique bar in Santa Fe that offers entertainment. However, the bar is small, and they enjoy only if at most $\Psi$ of the possible $N$ attendees are present, in which case they obtain a reward $r$ at a cost $0\leq c\leq r$ for going to the bar. Otherwise, they can stay home and watch stars with utility $0$. Players have two actions: \textmd{go} if they expect the attendance to be less than $\Psi$ people or \textmd{stay at home} if they expect the bar will be overcrowded.

The extension of the game introduces a learning component based on the belief of future attendance that every player has: the only information available is the number of people who came to El Farol in past weeks  \cite{Shang},\cite{Shang_2007},\cite{MarinaPetri200894} .

%

\f{{\em Incentive-based schemes for DTNs}}. Several incentive schemes have been recently proposed for DTNs. For example, \cite{shevade2008incentive} uses Tit-for-Tat (TFT) 
to design an incentive-aware routing protocol that
allows selfish DTN nodes to maintain hight performance
while satisfying TFT constraints. Mobicent \cite{chen2010mobicent} is an incentive-compatible system which uses credit with cryptographic method to incentive selfish nodes to forward packets. PI \cite{lu2010pi} attaches an
incentive on the sending bundle to stimulate the selfish nodes to cooperate in message delivery.  SMART \cite{zhu2009smart} is a secure multilayer credit-based incentive system for DTNs. In SMART, layered coins are used to provide incentives to selfish DTN nodes for bundle forwarding. MobiGame \cite{wei2011mobigame} is a user-centric and social-aware reputation based incentive scheme for DTNs.

\f{The remainder of this paper is organized as follows. The next section introduces the system model and the notation used throughout the paper. Results for the equilibria of the MG in device-dependent heterogeneous DTNs are derived in in Sec.~\ref{sec:multiclass}. The extension to the incomplete information scenario in user-dependent heterogeneous DTN is provided in Sec.~\ref{sec:user-dependent}. A distributed reinforcement learning algorithm able to drive the system to the desired operating point is derived in Sec.~\ref{sec:algo}. Numerical results for validating the outcomes of the theoretical analysis are reported and discussed in Sec.~\ref{sec:numerical}. Final remarks are reported in Sec.~\ref{sec:concl}.}


\section{Network model}\label{sec:netmodel}


In this section, we present the overall architecture and key insight into our mechanism design.

\subsection{System architecture and reward mechanism}

We consider a DTN with several source-destination pairs  $s_i$  and  $N$ relay nodes.  Relay nodes are equipped with a wireless interface allowing communication
with other mobiles in their proximity. Messages are generated at the source nodes and need to be delivered to the destination nodes. The network is assumed to
be sparse: nodes are isolated with high probability at any time instant.\footnote{This is also the case when disruption caused by mobility occurs at a fast pace compared to the typical operation time of protocols, e.g., the TPC/IP protocol suite.} Communication opportunities arise whenever two  nodes fall within reciprocal radio range, i.e., a ``contact'' occurs. We assume contacts last enough to ensure the transmission of all data needed for a message relaying. Also, we assume that inter contact  times  between any pair of nodes are independent identically distributed (i.i.d.) random variables. Note that to this respect our framework generalizes i.i.d. models proposed in literature to mimic synthetic mobility processes, e.g., Random Walk or Random Waypoint, since we allow for general intermeeting distributions, which may well include heavy tailed ones.

A source attempts  to deliver a message to destination generating several copies among  relays. Each such copy contains a time stamp indicating its age and can be dropped when it becomes irrelevant, e.g., after time $\tau$. $\tau$ is also the horizon by which we intend to optimize network performance. Due to lack of permanent connectivity, we exclude the use of feedback that allows the relays or sources to know whether the message has been successfully delivered to its destination or not. For the same reason, the design of our activation mechanism should not require centralized coordination \fdp{nor full state information} and any such scheme should indeed run fully distributed on board of the relay nodes.

Let  $g_{j}$ be the   energy cost for relay node $j$ when it remains active during $[0,\tau]$. This cost captures the heterogeneity of nodes in DTN in terms of cost for activation. 

Now assume that aim is to achieve a target performance figure (e.g., delivery probability, end-to-end delay).  Without loss of generality, we focus on the probability of  successful delivery. However,  our results extend to any performance measure monotonically increasing with the number of active relays. Given target performance figure and the parameters of the DTN (e.g., mobility, transmission range,  density of nodes), it is possible to estimate the number of nodes which should activate, named $\Psi$, in order to guarantee this target figure.  Now the question is how to stimulate $\Psi$ user nodes to participate to delivery message in a distributed manner.

To this aim each source $s$ proposes a reward for relays. For instance, the reward can be a certain number of credits that relays may use to send their own messages over the DTN. Also, the reward  $r^s_j$ is based on the type of device $j$. In fact, nodes  with larger battery capacity might choose to be more active to collect the reward, while nodes with a limited battery capacity may participate less to save energy.\footnote{We assume that the source is able to distinguish between classes according to node's type (whether it is a throwbox, a smartphone, etc) and determine the reward for each node according its type.}  In particular, the relay node of type $j$ receives a positive reward $r^s_j$ if and only if it is the first one to deliver the message to the corresponding destination.

Overall, sources satisfy performance requirements by activating relays by rewarding:  larger rewards engender more nodes to be active which yields higher delivery probability at the expense of battery depletion and network's lifetime. This trade-off rises the following question: {\em How to define the reward in order to activate enough relay nodes such as  to  attain the assigned  performance figure?}

The answer to this question is investigated hereafter.

\subsection{Network Game}

When a message is generated by a source node, competition is engendered by the general incentive mechanism during the message lifetime $\tau$. Each mobile has two strategies: either to participate to forwarding, i.e., pure strategy {\em transmit} $(T)$, or not to participate, i.e., pure strategy {\em silent} $(S)$.
Each strategy corresponds to a certain utility for the relay. Clearly, the payoff of a relay should depend on the actions performed by $N-1$ opponent mobiles. In our mechanism, the utility is designed such in a way that, for each player in the game, it is worth playing a given action if the number of peer nodes that adopt the same strategy does not exceed  $\Psi$, i.e., fraction of the total population of interacting nodes. In fact, mobiles of each class $j$ who take the minority action, within the tagged class, win, whereas the majority loses. To this respect activation threshold $\Psi$ is the minority rule of our game.

Let's now detail how the minority game develops.  Assume target probability of successful delivery $D_{succ}^{s}$:
\begin{equation}
D_{succ}^{s}\geq D_{succ}^{th}.
\label{target}
\end{equation}
From the sources' point of view,  the probability of successful delivery of a message is given by
\bear \label{eq:Psource}
D_{succ}^{s}(|{\mathcal N}_T|)&&=1-\prod_{k\in {\mathcal N}_T} Q_{\tau}^k,
\eear
where  ${\mathcal N}_T$  is the set of active relay nodes and $Q^k_{\tau}$ is the probability that  node $k$ fails to relay the copy of the message to the destination. Then $1 - Q^k_{\tau}$ is the probability that the tagged node succeeds to rely the copy of the message to the destination within time $\tau$. The expression of $Q_{\tau}^k$ depends on the distribution of the inter-meeting intervals.


The number $\Psi$ of active relays will then define such as that target performance figure $D_{succ}^{s}(\Psi)=D_{succ}^{th}$ .

%

Formally, let $k_T $ ($k_S$) be   the number of agents selecting strategy $T$ (resp.  $S$). A tagged relay playing strategy $T$ is member of the minority if $k_{T} \leq \Psi$, otherwise it loses; silent agents win as $k_S \leq N - \Psi$.    Hence the total reward of an active relay in class $j$ is given by  $R_j = \sum_{s}r_j^s P_{succ}^s(T, k,s)$, \f{where $P_{succ}^s(T, k,s)$ is the probability that an active node (of class $j$) receives reward $r_j^s$ from source $s$ when $k$ nodes are active}.  For the sake of simplicity and clearness we assume that every node has the same probability to meet a source and thus all sources  use the same  mechanism  reward, i.e., $r_j^s=r_j$. Hence the total reward of a relay  of type $j$ becomes  $R = n_s r_j P_{succ}(T, k,s)$, where $n_s$ is the number of sources in the network.

In the rest of the paper, we make a key assumption on function $P_{succ}(T, k,s)$ which follows naturally since only the first relay to deliver obtains the reward:

\f{{\bf Assumption A}: $P_{succ}(T, k,s)$ is decreasing in the number of active relays $k$.}

%
%
%
%
%



\section{Homogeneous Energy cost}
In this section we study a homogeneous DTN (in terms of energy cost). We consider all nodes have the same energy cost $g_j=g, \forall j$.
Now we can introduce two utility functions for our game, under the assumption that the population of sources is homogeneous: $P^{s}_{succ}(T, k,s) = P_{succ}(T, k)\; \forall s$:

\paragraph*{\bf Scenario 1} Zero-sum utility
$$
\vspace{-6mm}
\hskip-3mmU(T,k_T) = \sum_s r^s  \cdot P_{succ}^s(T,k_T,s) - g\tau,\;  U(S,k_S) = -U(T,k_T)
$$
\paragraph*{\bf Scenario 2} Fixed regret utility
$$
\vspace{-2mm}
U(T,k_T) =  \sum_s r^s  \cdot P_{succ}^s(T,k_T,s) - g\tau, \quad U(S,k_S) = -\alpha, \forall \;k_S
$$
where in the second case the utility of non-active nodes expresses the regret or satisfaction
for not participating to message relaying. In particular, we assume $\alpha \geq  0$, and we define
$k_T^{\alpha}$ such that $U(T,k_T^{\alpha}) = -\alpha$.

The formulation of {\bf Scenario 1}, requires nodes to estimate $P_{succ}$. This can be calculated over time by interrogating neighboring nodes and averaging their success rate: this amounts to run a pairwise averaging protocol as in \cite{Guer2010}. In case we want to avoid the use of gossip mechanisms, we can model regret of non-active nodes as a constant negative perceived utility, which corresponds to {\bf Scenario 2}.


Later on, we provide the exact characterization of the equilibria induced by the game: we distinguish pure Nash equilibria  and mixed Nash equilibria.

 \subsection{\bf Pure Nash Equilibrium}

The Nash Equilibrium in pure strategy for our game is given by the relation :
 \begin{eqnarray}
 &U(S, k_T)\geq  U(T, k_T+1) \mbox{ and } U(S, k_T-1)\leq U(T, k_T)\label{eq:nashcondS}
 \end{eqnarray}
 Thus, no player can improve its utility by unilaterally deviating from the equilibrium.

 \begin{proposition} \label{prop:pure}
 Under assumption A, there exists a pure Nash Equilibrium for our game. Moreover \\
 (i) for {\bf scenario 1},  there exists a unique NE obtained when exactly $\Psi$ among the total population of $N$ nodes play $T$.\\
 (ii)  for {\bf scenario 2}, there  exists two Nash equilibria which are obtained when the total number of active relays is such that: $k_T\in\{k_T^{\alpha},k_T^{\alpha}-1\}$
 \end{proposition}

 \Prf
\noindent{Scenario 1}: First, we show that $k_T=\Psi$ is a pure Nash equilibrium:
 $$
 U(S,\Psi) = U(T,\Psi) = 0 \geq U(T,\Psi +1).
 $$
 which is first condition (\ref{eq:nashcondS}-1). In the same way
 $$
 U(S,\Psi -1) = -U(T, \Psi -1) \leq 0 = U(T,\Psi)
 $$
 and we have second condition (\ref{eq:nashcondS}-2).

Second, we show that at the NE: $(k_T, k_S) = (\Psi, N - \Psi)$. By contradiction: let $k_T> \Psi \Rightarrow U(S, k_T) \geq  U(T, k_T+1)$, i.e., (\ref{eq:nashcondS}-1) holds. However, $$U(S, k_T -1) = - U(T, k_T - 1) \geq 0 > U(T, k_T)$$ and (\ref{eq:nashcondS}-2) fails. Conversely, let $k_T < \Psi \Rightarrow  U(S, k_T-1) \leq U(T, k_T)$ so that (\ref{eq:nashcondS}-2) holds. But, $$U(S,k_T) = -U(T,k_T)< 0 \leq U(T,k_T+1)$$ and (\ref{eq:nashcondS}-1) fails. Hence, $k_T=\Psi$ is the only possible pure Nash equilibrium.

\noindent{Scenario 2}: Let $k_T\in\{k_T^{\alpha},k_T^{\alpha}-1\}$ we have,
\[
\left\{
\begin{array}{cccc}
U(S,k_T)=-\alpha=U(T,k_T^{\alpha})&\geq& U(T,k_T+1), \\
U(S,k_T-1)=-\alpha=U(T,k_T^{\alpha})&\leq&U(T,k_T),
\end{array}
\right.
\]
where equality holds in the first relation for $k_T = k_T^{\alpha}-1$ and in the second for $k_T = k_T^{\alpha}$. We show that if $k_T\not\in\{k_T^{\alpha},k_T^{\alpha}-1\}$ then $(k_T,N-k_T)$ then (\ref{eq:nashcondS}-1) or (\ref{eq:nashcondS}-2) fails. In fact, if $k_T>k_T^{\alpha}$ we have,
\begin{eqnarray}
U(S,k_T)=-\alpha=U(T,k_T^{\alpha})\geq U(T,k_T+1), \mbox{but:}\nonumber \\
U(S,k_T-1)=-\alpha=U(T,k_T^{\alpha}) > U(T,k_T)\nonumber
\end{eqnarray}
Second, if $k_T<k_T^{\alpha}-1$ we have,
\begin{eqnarray}
U(S,k_T-1)=-\alpha=U(T,k_T^{\alpha})< U(T,k_T), \mbox{but:}\nonumber\\
U(S,k_T)=-\alpha=U(T,k_T^{\alpha})<U(T,k_T+1)\nonumber
\end{eqnarray}
Which concludes the proof for the second scenario.
\endpf
\begin{remark}
A crucial design issue is how to relate the parameters of the game to the performance of the DTN at the equilibrium.
From (\ref{eq:Psource}), the number of active nodes required to attain $D_{succ}^{th}$ needs to verify $k_T^{th}(k_T^{th}+1)=\frac{2\log(1-D_{succ}^{th})}{\log(Q_{\tau})}$. Besides, from Proposition \ref{prop:pure} it must be $\Psi=k_{T}^{th}$. We obtain: \[r^{*}=g\tau\frac{1}{n_sP_{succ}(T,k_{T}^{th})}\]
Message reward $r$ at the equilibrium is proportional to energy cost $g$ through a positive constant.
\end{remark}

\subsection{\bf Mixed Nash Equilibrium}

Let's consider now that relay nodes maintain a probability distribution over the two actions. Compared to pure strategy game, in the mixed strategy game every node can define the strategy by which it will be active only for a fraction of the time and stay silent the rest of the time. This kind of equilibrium is desirable for an homogeneous population of nodes with similar energy constraints.

In the mixed strategy game, node $i$ can choose to play action $T$ with probability $p_i$ and play $S$ with probability $(1-p_i)$. We let, ${\bf p}=(p_1,p_2,...,p_N)$, $p_i\geq0,\;\forall i$ the mixed strategy profile of our game. If $0<p_i<1,\;\forall i$ then ${\bf p}$ is a fully mixed strategy profile of the game.  A standard companion notation that we use for ${\bf p}$ is $(p_i,{\bf p_{-i}})$: it denotes the strategy profile of the game when relay $i$ uses strategy $p_i$ and others use ${\bf p}_{-i}= (p_1,..,p_{i-1},p_{i+1},..,p_N)$. Let's denote by $U_i(\tilde p, {\bf p}_{-i} ) $ the utility of node $i$ playing action $T$  with probability $\tilde p$. We have the following definition of the mixed strategy Nash Equilibrium:
\begin{definition}
\begin{itemize}
\item[(i)]A {\bf \em mixed strategy Nash Equilibrium} specifies a mixed strategy $p^*_i \in [0,1]$ for each player $i$ $(where \; i=1\ldots N)$ such that :
\beq
U_i(p^*_1,..,p^*_{i-1},p^*_i ,p^*_{i+1},..,p^*_N ) \geq U_i(p^*_1,..,p^*_{i-1},p_i ,p^*_{i+1},..,p^*_N )
\label{eq:cond_mixed}
\eeq
for every mixed strategy $p_i \in  [0,1]$.\\
\item[(ii)] We call a {\bf \em Fully mixed Nash Equilibrium} a mixed strategy Nash equilibrium ${\bf p}= (p_1,..,p_{i},..,p_N)$ with $p_i \not\in \{0,1\}, \forall i$.
\end{itemize}
\end{definition}
From now on we will denote by the term {\bf 'mixer'} a relay who uses a mixed strategy $0< p_i < 1$.
The following proposition states that any mixed equilibrium ${\bf p}$ with $p_i \not\in \{0,1\} \forall i$, is symmetric, i.e.  $p_i = p\; \forall i$. This result comes from the fact that given any pair of mixers, a player is better off if the other chooses differently. Moreover, at the equilibrium each player must be indifferent on whether it is active or silent.

\begin{proposition}\label{sym}

Assume assumption A holds.  Let ${\bf p}$ be the  mixed strategy profile of our game $s.t \; p_i \not\in \{0,1\}$, then at the equilibrium, all mixers must use the same probability $p$, i.e.,    $p_i = p_j\; \forall \mbox{ mixer } i,j$.
\end{proposition}

\Prf
Assume that the set of mixers is not empty and let suppose that there are $l$  relays that select pure strategy  $T$ and $r$ pure strategy $S$. Without loss of generality let the strategy profile at the equilibrium :
$${\bf p} = (p_1,\ldots,p_{N-l-r},1,\ldots,1,0,\ldots,0)$$
\noindent{{\bf Scenario 1}}: The utility for a mixer relay $i$ writes 
$$
U_i(\tilde p, {\bf p}_{-i} ) = (2\tilde p_i - 1)W(p_1,p_2,\ldots,p_{i-1},p_{i+1},\ldots,p_N)
$$
with
\begin{eqnarray*}
W(p_1,p_2,\ldots,p_{i-1},p_{i+1},\ldots,p_N)&=&\overset{N-l-r}{\underset{j \neq i}{\prod}}(1 - p_j)U(T,l+1) + \overset{N-l-r}{\underset{j \neq i}{\sum}}p_j\overset{N-l-r}{\underset{j' \not\in \{i,j\}}{\prod}}(1-p_{j'})U(T,l+2)+\\
&&\hskip-5mm  \overset{N-l-r}{\underset{j,j' \neq i}{\sum}}p_jp_{j'}\overset{N-l-r}{\underset{j'' \not\in \{i,j,j'\}}{\prod}}(1-p_{j''}) U(T,l+3)+ ... + \overset{N-l-r}{\underset{j \neq i}{\prod}}p_j U(T,N-r).
\end{eqnarray*}
Note about this function that:
\begin{itemize}
\item $W$ is strictly decreasing by any unilateral increase of $p_j$ by node $j$. This comes from the fact that the utility function of an active node is decreasing with the number of active nodes (assumption A).
\item For any two mixers $j \neq j'$, $p_j$ and $p_{j'}$ are indifferently  interchangeable variables in $W$.
\end{itemize}

At mixed equilibrium ${\bf p}$,  $\frac{\partial U_i({\bf p})}{\partial \tilde p_i} = 0\; \forall\; i\in\{1,\ldots,N-l-r\}$. This implies that: $$W(p_1,p_2,\ldots,p_{i-1},p_{i+1},\ldots,p_N) = 0,\forall  \mbox{ mixer }i$$. Now suppose that there exists two mixers $i$ and $j$, s.t. $p^*_i \neq p^*_j$. Without lost of generality assume that $ p^*_i < p^*_j$, then
\begin{eqnarray*}
0 = W(p,..,p_{i-1},p_{i+1},..,p_j,..,p_N) > W(p_1,..,p_{i-1},p_{i+1},..,p_i,..,p_N)\\
= W(p_1,.., p_{j-1},p_{j+1},..,p_N) >  0
\end{eqnarray*}
which is absurd. Thus $p_i = p_j,\,\forall \mbox{ mixers }i,j$.

\noindent{{\bf Scenario 2}}: As in {\bf Scenario 1}, the utility perceived by a given mixer $i$ when the strategy profile is ${\bf p}= (p_1,p_2,\ldots,p_N)$ is given by:
\begin{eqnarray*}
\hskip-2mmU_i(\tilde p, {\bf p}_{-i}) &=& \tilde p_i W^{'}(p_{-i}) -\alpha(1 - \tilde p_i)\Big[\overset{N-l-r}{\underset{j \neq i}{\prod}}(1 - p_j)  + \overset{N-l-r}{\underset{j \neq i}{\sum}}p_j\overset{N-l-r}{\underset{j' \not\in \{i,j\}}{\prod}}(1-p_{j'}) \\
&&+ \overset{N-l-r}{\underset{j,j' \neq i}{\sum}}p_j p_{j'}\overset{N-l-r}{\underset{j'' \not\in \{i,j,j'\}}{\prod}}(1-p_{j''}) + ... + \overset{N-l-r}{\underset{j \neq i}{\prod}}p_j\Big]
\end{eqnarray*}
At the equilibrium we have, $\forall \mbox{ mixer  }i, \, \frac{\partial U_i({\bf P})}{\partial p_i}= W^{'}(p_{-i})= 0$, where $W^{'}$ has exactly the same shape as $W$ with $U(T,k)$ replaced by $U(T,k)+\alpha$, $k\in\{l+1,\ldots,N-r\}$. We then use the same reasoning as done with function $W$ and conclude that, $p^*_i = p^*_j,\,\forall \mbox{ mixers }i,j$. \endpf\\

In the following corollary, we restrain the result of proposition \ref{sym} to the special case when every nodes act as mixers. \\
\begin{corollary}
Under assumption A, any fully mixed equilibrium ${\bf p}$ with $p_i \not\in \{0,1\}, \forall i$, is symmetric, i.e.  $p_i = p\; \forall i$.\\
\end{corollary}
The following proposition  characterize the existence and uniqueness of a fully mixed Nash Equilibrium.\\

\begin{proposition}\label{prop:sym}
Under assumption A, there exists a unique fully mixed Nash Equilibrium ${\bf p}^*$. Moreover, ${\bf p}^*$ is solution to:
 \begin{itemize}
 \item {\bf Scenario 1 }:
\beq
\vspace{-1mm}
\label{eq:homo_equi1}
\hspace{-0.8cm}A(N,p^*)= \underset{k=1}{\overset{N}{\sum}} C_{k-1}^{N-1} (p^*)^{k-1}(1-p^*)^{N-k}U(T,k) = 0.
\eeq

\item  {\bf Scenario 2} :
\beq
\vspace{-1mm}
\label{eq:homo_equi2}
\hspace{-1cm}A'(N,p^*) = \underset{k=1}{\overset{N}{\sum}} C_{k-1}^{N-1} (p^*)^{k-1}(1-p^*)^{N-k}[U(T,k)+\alpha] = 0. \nonumber
\eeq
\end{itemize}
\end{proposition}
\Prf
Let $p$ the symmetric mixed strategy adopted by every node in the game, $p_i = p,\;\forall i$.

\noindent{{\bf Scenario 1}}: The utility of one relay $i$ when the strategy profile $(p_i,p_{-i})$ is played  is given by:
\begin{eqnarray}
U_i(\tilde p_i,p_{-i})&=& \tilde p_i\underset{k=1}{\overset{N}{\sum}} C_{k-1}^{N-1} p_{-i}^{k-1}(1-p_{-i})^{N-k}U(T,k) +  (1-\tilde p_i)\underset{k=0}{\overset{N-1}{\sum}} C_{k}^{N-1}p_{-i}^{k}(1-p_{-i})^{N-k-1}U(S,k+1) \nonumber\\
&=& \tilde p_i\underset{k=1}{\overset{N}{\sum}} C_{k-1}^{N-1} p_{-i}^{k-1}(1-p_{-i})^{N-k}U(T,k) + (1-\tilde p_i)\underset{k=1}{\overset{N}{\sum}} C_{k-1}^{N-1}p_{-i}^{k-1}(1-p_{-i})^{N-k}U(S,k) \nonumber\\
&=& (2\tilde  p_i-1)\underset{k=1}{\overset{N}{\sum}} C_{k-1}^{N-1} p_{-i}^{k-1}(1-p_{-i})^{N-k}U(T,k) \nonumber
\end{eqnarray}

   Let $A(N,p_{-i}) = \underset{k=1}{\overset{N}{\sum}} C_{k-1}^{N-1} p_{-i}^{k-1}(1-p_{-i})^{N-k}U(T,k)$\\
if $A(N,p_{-i}) < 0$, $p_i = 0$ is the best response for player $i$ and conversely, $p=1$ is a best response when $A(N,p_{-i}) >0$. A mixed strategy is obtained when $A(N,p_{-i}) = 0$. Also, we have $$A(N,0) = U(T,1) >0 >A(N,1) = U(T,N)$$ thus there exists a mixed symmetric Nash Equilibrium which is unique since  $A(N,p_{-i})$ is strictly decreasing with $p$. The mixed equilibrium is thus characterized by equation (\ref{eq:homo_equi1}).
$$A(N,p^*) = \underset{k=1}{\overset{N}{\sum}} C_{k-1}^{N-1} (p^*)^{k-1}(1-p^*)^{N-k}U(T,k) = 0.$$

\noindent{{\bf Scenario 2}}: The utility of one relay $i$ when the strategy profile $(\tilde p_i,p_{-i})$ is played  is given by:
\begin{eqnarray}
U_i(\tilde p_i,p_{-i}) &=& \tilde p_i\underset{k=1}{\overset{N}{\sum}} C_{k-1}^{N-1} p_{-i}^{k-1}(1-p_{-i})^{N-k}U(T,k) - \alpha(1-\tilde p_i)\nonumber
\end{eqnarray}

At the Nash equilibrium we have, $\forall \mbox{ player }i, \, \frac{\partial U_i(p^*)}{\partial p^*}=A'(N,p^*)=0$ with $$A'(N,p^*) = \underset{k=1}{\overset{N}{\sum}} C_{k-1}^{N-1} (p^*)^{k-1}(1-p^*)^{N-k}[U(T,k)+\alpha]$$

Since $\alpha$ is a fixed positive constant, $A'(N,p^*)$ has the same properties as $A(N,p^*)$ from the proof of scenario 1. Then we easily conclude that, $p^*$ is unique and characterized by :
$$A'(N,p^*) = \underset{k=1}{\overset{N}{\sum}} C_{k-1}^{N-1} (p^*)^{k-1}(1-p^*)^{N-k}[U(T,k)+\alpha] = 0.$$
\endpf

\subsection{\bf Equilibrium with Mixers and Non-mixers}
\label{sec:mixer}
We study here the existence of equilibrium when the population of agents is composed of pure strategy players: active or non-active, as well as mixers. In this case, a non-pure Nash equilibrium can be represented by the triplet $(l,r,p^*)$, where $l,r\in\{0, 1,\ldots,N\}$ denote respectively the number of agents choosing pure strategy $T$ or $S$, and $p^*\in(0,1)$ the probability with which the remaining $N-l-r$ mixers choose strategy $T$. Moreover, we denote by $v_T(l,r,p)$(resp. $v_S(l,r,p)$) the expected payoff to a player choosing $T$(resp. $S$). The expressions of $v_T(l,r,p) \mbox{ and }v_S(l,r,p)$ write as follow:
\begin{equation}\label{eq:vt}
v_T(l,r,p) = \underset{k=0}{\overset{N-l-r}{\sum}} C_{k}^{N-l-r}p^k(1-p)^{N-l-r-k}U(T,l+k)
\end{equation}
and
\begin{equation}\label{eq:vs}
v_S(l,r,p) = -\underset{k=0}{\overset{N-l-r}{\sum}} C_{k}^{N-l-r}p^k(1-p)^{N-l-r-k}U(T,l+k)
\end{equation}

\begin{proposition}
Using the previous notations, a strategy profile of type $(l,r,p^*)$ is a Nash equilibrium with at least one mixer if and only if:
\begin{equation}\label{eq:NashEqCond}
v_T(l+1,r,p^*) = v_S(l,r+1,p^*)
\end{equation}
We prove that this result holds for  Zero-sum utility and fixed regret utility for non-active nodes(resp. {\bf Scenario 1} and {\bf Scenario 1})
\end{proposition}

\Prf
The condition (\ref{eq:NashEqCond}) describes that a mixer is indifferent whether it chooses a pure strategy $T$ or $S$. This is a necessary condition for the strategy profile $(l,r,p^*)$ to be a Nash equilibrium.\\
 In order to show sufficiency, we need to show that pure strategy players as well, cannot improve their expected utility through unilateral deviation from the equilibrium profile.
Without loss of generality, suppose that there is at least one player using pure strategy $T$, we have
\begin{eqnarray}\label{eq:prfNash}
v_T(l,r,p^*) &\geq& v_T(l+1,r,p^*)=v_S(l,r+1,p^*)\nonumber\\
&\geq& v_S(l-1,r+1,p^*)\nonumber\\
&\geq& p^*v_T(l,r,p^*) +  (1-p^*)v_S(l-1,r+1,p^*)\nonumber
\end{eqnarray}

This last relation, states that an active user cannot improve its expected utility by unilaterally deviating from the strategy profile $(l,r,p^*)$ using any strategy $p^*\in [0,1)$, given relation $(\ref{eq:NashEqCond})$.
As done for {\bf Scenario 1}, in {\bf Scenario 1}, we have, $v_S(l,r+1,p^*) = -\alpha$, let $v_S(l+1,r,p^*)=-\alpha$ then:
\begin{eqnarray}\label{eq:prfNash2}
v_T(l,r,p^*) \geq& v_T(l+1,r,p^*)=-\alpha \geq v_S(l-1,r+1,p^*)\nonumber\\
\geq& p^*v_T(l,r,p^*) +  (1-p^*)v_S(l-1,r+1,p^*)\nonumber
\end{eqnarray}
moreover,
$$v_S(l+1,r-1,p^*) \leq v_T(l+1,r,p^*) = -\alpha =v_S(l,r,p^*).$$
This completes the proof.
 \endpf\\

 {\bf Discussion on existence of $(l,r,p^*)$ type equilibria}\\
 It is possible to isolate several cases where the relation (\ref{eq:NashEqCond}) that characterizes a Nash Equilibrium of type $(l,r,p^*)$, cannot be satisfied. \\
 We denote by, $p = 0^+$(resp. $p = 1^-$) the mixed strategy infinitely close to zero(resp. to one), with which at least one mixer select to be active.
 Since, $v_T(l,r,p^*)$ is strictly decreasing with $l$ and $p^*$, we have,
 $v_T(l+1,r,p^*) = v_S(l,r+1,p^*)$
\[
\iff \left\{
\begin{array}{cccc}
 v_T(l+1,r,0^+) &>& -v_T(l,r+1,0^+)\\
 v_T(l+1,r,1^-) &\leq& -v_T(l,r+1,1^-),
\end{array}
\right.
\]
 \begin{itemize}
 \item[(1)] If $l \geq \Psi$, then there is no Nash equilibrium of the desired type. Indeed, $l>\Psi$, then  $v_T(l,r+1,0^+) \leq 0$ and
$$v_T(l+1,r,0^+) \leq 0 \leq -v_T(l,r+1,0^+).$$ Then there is no possible Nash Equilibrium according to relation $(\ref{eq:NashEqCond})$.

\item[(2)] If $l+r+1 > N-1$, then there is no Nash equilibrium. We already have $l<\Psi$, let $l+r+1 = N$ then,
$$v_T(l+1,r,p) = C_1\geq 0 \;\forall\; p\mbox{ and } $$
$$ v_S(l,r+1,p)= C_2> 0 \;\forall\; p.$$
Since $v_T$ is decreasing with $l$, we have, $0\leq C_1< C_2$ which contradicts relation (\ref{eq:NashEqCond}).
 \end{itemize}
 A Nash Equilibrium of type $(l,r,p^*)$ exists then only for $l<\Psi$ and for $l+r\leq N-2$, thus there are exactly $\Psi(N-2)-\frac{\Psi(\Psi-1)}{2}$ Nash equilibria. In the following proposition we go further and decline some properties of the symmetric mixed strategy $p^*$ at the equilibrium.

 \begin{proposition}
 \label{prop:mixer}
 The mixed strategy $p^*$ at the equilibrium increases as $r$ increase and reversely decreases as $l$ increase.
 \end{proposition}

 \Prf
 For a fixed number $l$ of nodes playing pure strategy $T$, the utility of a mixer when there are less nodes playing pure strategy $S$, decreases faster than when there are  more nodes playing pure strategy $S$. For example we have,
 $$
 \frac{\partial v_T(l+1,0,p)}{\partial p} >  \frac{\partial v_T(l+1,1,p)}{\partial p}
 $$
 Similarly, we will have
  $$
 \frac{\partial v_T(l,1,p)}{\partial p} >  \frac{\partial v_T(l,2,p)}{\partial p}.
 $$
 Since, $v_T(l+1,0,0^+) = v_T(l+1,1,0^+)$ and $v_T(l,1,0^+) = v_T(l,2,0^+)$ then if $p_1^*,p_2^*$ are such that $v_T(l+1,0,p_1^*) = -v_T(l,1,p_1^*)$ and $v_T(l+1,1,p_2^*) = -v_T(l,2,p_2^*)$, it follows that $p_1^*<p_2^*$.\\
  The same reasoning holds for every $k<k' \mbox{ and } p_1^*,p_2^* \; s.t.\;v_T(l+1,k,p_1^*) = -v_T(l,k+1,p_1^*)$ and $v_T(l+1,k',p_2^*) = -v_T(l,k'+1,p_2^*)$ then $p_1^*<p_2^*$. \\
  We apply a similar reasoning reversely and conclude that  for a fixed number $r$ of nodes playing pure strategy $S$, for every $k<k' \mbox{ and } p_1^*,p_2^* \; s.t.\;v_T(k+1,r,p_1^*) = -v_T(k,r+1,p_1^*)$ and \\
  $v_T(k'+1,r,p_2^*) = -v_T(k',r+1,p_2^*)$ then $p_1^*>p_2^*$.\endpf\\

\paragraph*{{\bf Summary on characterization of equilibria}}
Throughout this section we have characterized the following different equilibria : Under assumption A we have
\begin{itemize}
\item[1.] {\em pure equilibrium} : We show that for the Zero-sum utility there exists a unique pure N.E. that sets at exactly $\Psi$ active relay nodes. For the Fixed regret utility scenario, there exists two possible N.E. for a number of active nodes $k_T\in\{k_T^{\alpha},k_T^{\alpha}-1\}$.
\item[2.] {\em fully mixed equilibrium} : For both scenarios we shown that any fully mixed equilibrium ${\bf p}$ with $p_i \not\in \{0,1\}, \forall i$, is symmetric. Moreover, the mixed N.E. of our game is unique and characterized by : $A(N,p^*) = \underset{k=1}{\overset{N}{\sum}} C_{k-1}^{N-1} (p^*)^{k-1}(1-p^*)^{N-k}U(T,k) = 0$ for the zero-sum utility scenario and characterized by $A'(N,p^*) = \underset{k=1}{\overset{N}{\sum}} C_{k-1}^{N-1} (p^*)^{k-1}(1-p^*)^{N-k}[U(T,k)+\alpha] = 0$ for the fixed regret scenario.
\item[3.] {\em equilibrium with mixers and non-mixers}: The last characterized type of equilibrium is related to a population of relays composed of mixers and non-mixers. Here we shown that such type of equilibrium is characterized by a specific relation, namely relation  (\ref{eq:NashEqCond}). Moreover, we established that a Nash Equilibrium of this type exists only for $l<\Psi$ and for $l+r\leq N-2$, thus there are exactly $\Psi(N-2)-\frac{\Psi(\Psi-1)}{2}$ Nash equilibria.
 \end{itemize}

\section{Heterogeneous  devices-dependent energy cost } \label{sec:multiclass}


This section considers the case when the energy cost depends on the class of devices the relays belongs to, let them be, e.g., notebook computers,  ebook readers and tablet computers such as the iPad, netbooks or smartphones.

We consider $M$ classes of devices in the DTN.  Each class contains $N_j$ relays with $N=\sum_{j=1}^N N_j$. For the sake of clarity, we will often refer to the case $M=2$. However, all results can be easily extended to hold in general case.  Assume that $g_1>g_2$, i.e., nodes belong to class $1$ have higher energy cost than class $2$ when active. For example mostly devices such as smartphones have smaller power budgets compared to laptops: in turn, their energy cost is higher. Furthermore the power consumption of the WiFi radio represents a large fraction of the overall power consumed by a small device. Hence, we define the utility of an active  relay of class $j$
 $$
U_j(T,k_T) =   n_s r_j P_{succ}(T,k_T) - g_j\tau
$$

where $r_j$ is the reward  designed by sources for class $j$. The utility for a silent node is $U_j(S,k_T) =  0$.  
Now, we should  characterize the number of active relay in each class $j$, named $\Psi_j$, able to guarantee the target (\ref{target}). $\Psi_j$ is thus the per-class minority of our game for the multi-class scenario. This threshold can be achieved using the reward mechanism ${\bf r^*}=(r^*_j)_{j=1,..,M}$ which should obey the following relation
\beq
\vspace{-1mm}
\label{eq:equilibre}
\forall 1\leq j \leq M:\quad n_s r^*_j P_{succ}(T,\Psi_j)=g_j\tau
\eeq

\subsection{\bf Pure Nash Equilibrium}

\begin{definition}
A Nash Equilibrium in pure strategies exists if and only if the following two conditions to be satisfied:
\begin{equation}\label{eq:nashcond_hete}
\forall 1\leq j \leq M:\left \{
 \begin{array}{ccc}
 U_j(S, k_T) &\geq&  U_j(T, k_T+1) \\
 U_j(S, k_T-1) &\leq& U_j(T, k_T)
 \end{array}
\right.
 \end{equation}
 \end{definition}
 Actually, the above statement says that no player can improve its utility by unilaterally deviating from the equilibrium. The equilibrium of the multi-class games is as follow

 \begin{proposition} \label{pe_multi}
Let $ {\bf r}$ be the reward mechanism designed by sources. Then there exists a pure Nash equilibrium. Further, the number of active relay nodes for class $j$,  named $k_{T,j}$,  is  the same under all pure Nash equilibria where $k_{T,j}$ is solution of
 \beq
\forall 1\leq j \leq M:\quad r_j P_{succ}(T,k_{T,j})=g_j\tau
\label{rewa}
\eeq
\end{proposition}

\Prf

 Assume that for any class $j$ exactly $k_{T,j}$ nodes are active, then we have:
  \[
\left\{
\begin{array}{cccc}
U_j(S,k_{T,j})=0=U_j(T,k_{T,j})&\geq& U_j(T,k_{T,j}+1), \\
U_j(S,k_{T,j}-1)=0=U_j(T,k_{T,j})&\leq&U_j(T,k_{T,j}),
\end{array}
\right.
\]
then we have the conditions in (\ref{eq:nashcond_hete}) satisfied.

We now show that there are no other pure Nash equilibria. Let, for a class $j$, $k_{T,j}^{'}\neq k_{T,j}$, without loss of generality, let $k_{T,j}^{'}> k_{T,j}$ then
\begin{eqnarray}
U_j(S,k_{T,j}^{'})=0=U_j(T,k_{T,j})\geq U_j(T,k_{T,j}^{'}+1), \mbox{but:}\nonumber \\
U_j(S,k_{T,j}^{'}-1)=0=U_j(T,k_{T,j}) > U_j(T,k_{T,j}^{'})\nonumber
\end{eqnarray}
Second, if $k_{T,j}^{'}<k_{T,j}$ we have,
\begin{eqnarray}
U_j(S,k_{T,j}^{'}-1)=0=U_j(T,k_{T,j})< U_j(T,k_{T,j}^{'}), \mbox{but:}\nonumber\\
U_j(S,k_{T,j}^{'})=0=U_j(T,k_{T,j})<U_j(T,k_{T,j}^{'}+1)\nonumber
\end{eqnarray} and the second relation is not satisfied. Therefore at the equilibrium there are exactly $k_{T,j}$ active nodes. Hence the proof.
\endpf

We note that under reward mechanism $r$ in ({\ref{rewa}), there are  $\sum_{j=1}^{j=M} \Big( \begin{array}{c} N_j\\ k_{T,j} \end{array}\Big)$ pure Nash equilbria.  Further, under reward mechanism $r^*$ defined in (\ref{eq:equilibre}), the number of active relay nodes  in each class at a pure Nash equilibrium, is $\Psi_j$ with  $\Psi=\sum_{j=1}^{j=M}\Psi_j$ . This corresponds to the target of sources in the system.  Unfortunately the pure equilibrium  could fail  to achieve  a certain fairness between node relays since only a part of relay nodes participate  to relaying messages. To overcome this problem,  we use another concept of equilibrium, named full mixed equilibrium, in which a relay node  will be active only for a fraction of the time.   We also design a learning algorithm that allows the system to converge to a mixed equilibrium.

\subsection{\bf Mixed Nash Equilibrium}

Let's consider now that relay nodes maintain a probability distribution over the two actions. Compared to the pure strategy game, in the mixed strategy game every node can define the strategy by which it will be active only for a fraction of the time and stay silent the rest of the time.

In the mixed strategy game, node $i$ of class $j$ can choose to play action $T$ with probability $p_{ij}$ and play $S$ with probability $(1-p_{ij})$. Let the profile of our game in this multi-class framework ${\bf p} = (p_{11},...,p_{N_11},...,p_{1j},...p_{N_ij},...,p_{1M},...,p_{N_MM})$. If $0<p_{ij}<1,\;\forall i,j$ then ${\bf p}$ is a fully mixed strategy profile of the game.   We denote by $(p_{ij},{\bf p}_{-i})$ the fully mixed strategy profile of the game when relay $i$ of class $j$ uses strategy $p_{ij}$ and others use $p_{-i} = (p_{11},..,p_{N_11},..,p_{1j},..,p_{i-1j},p_{i+1j},..,p_{N_jj},..,p_{1M},..,p_{N_MM})$.
We have the following definition of the mixed strategy Nash Equilibrium:
\begin{definition}
\begin{itemize}
\item[(i)]A {\bf \em mixed strategy Nash Equilibrium} specifies a mixed strategy $p^*_{ij} \in [0,1]$ for each player $i$ $(where \; i=1\ldots N)$ such that :
  \bear
U^i(p^*_1,..,p^*_{i-1},p^*_i ,p^*_{i+1},..,p^*_N ) \geq \nonumber \\
U^i(p^*_1,..,p^*_{i-1},p_i ,p^*_{i+1},..,p^*_N ), \; \forall p_i
\label{eq:cond_mixed}
\eear
\item[(ii)] We call a {\bf \em Fully mixed Nash Equilibrium} a mixed strategy Nash equilibrium ${\bf p}$ with $p_i \not\in \{0,1\}, \forall i$.
\end{itemize}

\end{definition}

 \begin{proposition}\label{sym_hete}
 For any reward mechanism, at the mixed equilibrium, all players of the same class use the same  probability: $p_{ij} = p_j,\;\forall i; \forall 1\leq j \leq M$.
\end{proposition}

\Prf

The utility perceived by a given player $i$ of class $j$ when the strategy profile is $P$ is given by:
$$
U^i_j({\bf p}) = p_iW_i(p_{-i})
$$
with
\begin{eqnarray*}\label{func_f}
 W_i =&&  \underset{k\neq i}{\prod}(1-p_k)U_j(T,1)+\underset{k\neq i}{\sum}p_{km}\underset{k'\not\in \{i,k\}}{\prod}(1-p_{k'm})U_j(T,2) + \underset{k,k' \neq i}{\sum}p_{km}p_{k'm}\underset{k'' \not\in \{i,k,k'\}}{\prod}(1-p_{k''m}) U_j(T,3) \nonumber\\
&+& ... + \underset{k \neq i}{\prod}p_{km}U_j(T,N)
\end{eqnarray*}
$\forall 1\leq m \leq M$.
Note about this function that:
\begin{itemize}
\item $W_i$ is strictly decreasing by any unilateral increase of $p_{km}$ by player $k$ of class $m$.
\item For any two $k \neq k'$ of the same class $m$, the mixed strategies $p_{km},p_{k'm}$ are indifferently  interchangeable variables in $W_i$.
\end{itemize}

At the equilibrium we have, $\forall \mbox{ player }i, \forall 1\leq j \leq M,\, \frac{\partial U^i_j({\bf p})}{\partial p_{ij}} = 0$. This implies that : $W_i = 0$. Moreover, the strategy profile ${\bf p} = (p^*_{11},..,p^*_{N_11},..,p^*_{1j},..,p^*_{N_jj},..,p^*_{1M},..,p^*_{N_MM})$ is a Nash equilibrium if no user can increase its utility by any unilateral deviation.
Now suppose that there exists $i,k$ of class $j$, such that, $p^*_{ij} \neq p^*_{kj}$. Without lost of generality assume that $ p^*_{ij} < p^*_{kj}$, we have,
\begin{eqnarray*}
0 &=& W_i(...,p^*_{1j},...,p^*_{i-1j},p^*_{i+1j},...,p^*_{kj},...,p_{N^*_jj},...,p^*_{N_MM}) \\
&>& W_i(...,p^*_{1j},...,p^*_{i-1j},p^*_{i+1j},...,p^*_{ij},...,p_{N^*_jj},...,p^*_{N_MM})\\
&=& W_i(...,p^*_{1j},...,p^*_{k-1j},p^*_{k+1j},...,p_{N^*_jj},...,p^*_{N_MM})\\
&>& 0
\end{eqnarray*}
which is absurd. Thus $ p^*_{ij} = p^*_{kj},\,\forall\;i,k$ of class $j$.
\endpf

 Let $p_j$ be the symmetric mixed strategy adopted by every node of class $j$, $p_{ij} = p_j,\;\forall i,j$.
For reasons of clarity, we characterize the mixed strategy $p^*_j$ in a two-class scenario without any loss of generality (consider $M=2$).

\begin{proposition}\label{mixed_equi_multi_case}
Let $ {\bf r}$ be the reward mechanism designed by sources. Then there exists a unique fully mixed Nash equilibrium $(p_1^*,p_2^*)$ for the multi-class case. Moreover it is the solution of, $A_1(N,p_1^*,p_2^*)=A_2(N,p_1^*,p_2^*)=0$ where:
\begin{eqnarray}
\label{eq:a1a2}
\vspace{-10mm}
A_1(N,p_1^*,p_2^*) &=&\underset{k_1=0}{\overset{N_1-1}{\sum}}\underset{k_2=0}{\overset{N_2}{\sum}}( C_{k_1}^{N_1-1} p_{1}^{*k_1}(1-p_{1}^*)^{N_1-k_1-1}\nonumber\\
& & C_{k_2}^{N_2} p_{2}^{*k_2}(1-p_{2}^*)^{N_2-k_2})U_1(T,k_1+k_2)\nonumber\\
and \nonumber\\
\vspace{-10mm}
A_2(N,p_1^*,p_2^*) &=&\underset{k_1=0}{\overset{N_1}{\sum}}\underset{k_2=0}{\overset{N_2-1}{\sum}}( C_{k_2}^{N_2-1} p_{2}^{*k_2}(1-p_{2}^*)^{N_2-k_2-1}\nonumber\\
& & C_{k_1}^{N_1} p_{1}^{*k_1}(1-p_{1}^*)^{N_1-k_1})U_2(T,k_1+k_2)\nonumber
\end{eqnarray}
Moreover,
\begin{itemize}
\item[(i)] if $\frac{g_1}{r_1} = \frac{g_2}{r_2}$ then we have $p_1=p_2$.
\item[(ii)] if $\frac{r_1}{g_1}<\frac{r_2}{g_2}$ then we have  $p_1<p_2$. As a consequence  $k_{T,1}<k_{T,2}$.
\end{itemize}
\end{proposition}

\Prf

The utility of an active user of Class $1$ is given by:
\begin{eqnarray*}
U^i_1(p_{i1},p_{-i})=p_i\underset{k_1=0}{\overset{N_1-1}{\sum}}\underset{k_2=0}{\overset{N_2}{\sum}}( C_{k_1}^{N_1-1} p_{1}^{k_1}(1-p_{1})^{N_1-k_1-1}\nonumber\\
C_{k_2}^{N_2} p_{2}^{k_2}(1-p_{2})^{N_2-k_2})U_1(T,k_1+k_2)+ (1-p_i)*0\nonumber\\
=p_i\underset{k_1=0}{\overset{N_1-1}{\sum}}\underset{k_2=0}{\overset{N_2}{\sum}}\big( C_{k_1}^{N_1-1} p_{1}^{k_1}(1-p_{1})^{N_1-k_1-1}\nonumber\\
 C_{k_2}^{N_2} p_{2}^{k_2}(1-p_{2})^{N_2-k_2}\big)[U_1(T,k_1+k_2)] \nonumber\\
=p_i*A_1(N,p_1,p_2)\hspace{3cm} \nonumber
\end{eqnarray*}
and utility of user $i$ from Class 2 writes
\begin{eqnarray*}
U^i_2(p_{i2},p_{-i})= p_i\underset{k_1=0}{\overset{N_1}{\sum}}\underset{k_2=0}{\overset{N_2-1}{\sum}}( C_{k_2}^{N_2-1} p_{2}^{k_2}(1-p_{2})^{N_2-k_2-1}\nonumber\\
 C_{k_1}^{N_1} p_{1}^{k_1}(1-p_{1})^{N_1-k_1})[U_2(T,k_1+k_2)] + (1-p_i)*0 \nonumber\\
=p_i*A_2(N,p_1,p_2)\hspace{3cm} \nonumber
\end{eqnarray*}
where $A_1(N,p_1,p_2),A_2(N,p_1,p_2)$ are defined as follows:
\begin{eqnarray*}
A_1(N,p_1,p_2) &=&\underset{k_1=0}{\overset{N_1-1}{\sum}}\underset{k_2=0}{\overset{N_2}{\sum}}( C_{k_1}^{N_1-1} p_{1}^{k_1}(1-p_{1})^{N_1-k_1-1}\nonumber\\
& &  C_{k_2}^{N_2} p_{2}^{k_2}(1-p_{2})^{N_2-k_2})U_1(T,k_1+k_2),
\end{eqnarray*}
and
\begin{eqnarray*}
A_2(N,p_1,p_2) &=&\underset{k_1=0}{\overset{N_1}{\sum}}\underset{k_2=0}{\overset{N_2-1}{\sum}}( C_{k_2}^{N_2-1} p_{2}^{k_2}(1-p_{2})^{N_2-k_2-1} \nonumber\\
& & C_{k_1}^{N_1} p_{1}^{k_1}(1-p_{1})^{N_1-k_1})U_2(T,k_1+k_2).
\end{eqnarray*}

At the Nash equilibrium we have, $\forall \mbox{ player }i \mbox{ of class }j=1,2, \, \frac{\partial U^i_j(p^*)}{\partial p^*}=A_j(N,p_1^*,p_2^*)=0$,
if $A_j(N,p_1^*,p_2^*) < 0$, $p = 0$ is the best response for the player $i$ of class $j$ and conversely, $p=1$ is a best response when $A_j(N,p_1^*,p_2^*) >0$. A mixed strategy is obtained when $A_j(N,p_1^*,p_2^*) = 0, \forall j \in{1,2}$. Also, we have $A_j(N,p_1^*,p_2^*)$ is a strictly decreasing function in both $p_1,p_2$ (Assumption A). Thus there exists a mixed Nash Equilibrium which is unique and characterized by the equation (\ref{eq:mixed_hete_eqi}).
\beq
\label{eq:mixed_hete_eqi}
A_1(N,p_1^*,p_2^*)=A_2(N,p_1^*,p_2^*)=0.
\eeq

 Now let
\beq\label{eq:participate}
\hskip-3mmC(i) = \underset{k_1=0}{\overset{N_1-2}{\sum}}\underset{k_2=0}{\overset{N_2-2}{\sum}}P(K_1=k_1,K_2=k_2)r_iP_{succ}(T,k_1+k_2+e_i+1)\nonumber
\eeq
for user $i$, where $e_i=1$ if user $i$ is active and $e_i=0$ otherwise. We can thus rewrite the expressions of $A_1(N,p_1^*,p_2^*)$ and $A_2(N,p_1^*,p_2^*)$ as follows:
\bear
A_1(N,p_1^*,p_2^*)= r_1p_2C(1) - r_1(1-p_2)C(0) - g_1\tau\\
\vspace{-5mm}
A_2(N,p_1^*,p_2^*)= r_2p_1C(1) - r_2(1-p_1)C(0) - g_2\tau
\vspace{-4mm}
\eear

It follows that, $A_1(N,p_1^*,p_2^*)=A_2(N,p_1^*,p_2^*)=0\;\implies$
\begin{eqnarray}\label{eq:mixed_ed_cond}
p_2C(1) - (1-p_2)C(0) &=& \frac{g_1\tau}{r_1} \label{eq:mixed_ed_cond1}\\
p_1C(1) - (1-p_1)C(0) &=& \frac{g_2\tau}{r_2} \label{eq:mixed_ed_cond2}
\end{eqnarray}
letting $\frac{g_1\tau}{r_1} = \frac{g_2\tau}{r_2}$ we have, $p_1=p_2$. This completes the proof of $(i)$. \\
Now, let  $\gamma_1=\frac{g_1\tau}{r_1},\gamma_2=\frac{g_2\tau}{r_2}$ then
from  (\ref{eq:mixed_ed_cond1}) and (\ref{eq:mixed_ed_cond2}) we have:
\begin{eqnarray*}
(p_2-p_1)C(1) + (p_2-p_1)C(0)  &=& \gamma_1 -\gamma_2\\
\Rightarrow (p_2-p_1)(C(0) + C(1))  &=& \gamma_1 -\gamma_2
\end{eqnarray*}
Since, $C(0) > C(1)>0$\footnote{This comes from the fact that the more number of active nodes, the less is the probability of obtaining the reward for a tagged node.}
, then, $  \gamma_1 > \gamma_2 \Rightarrow p_2>p_1.$
This tells that in order to have fewer nodes active in class $1$ we should allocate smaller reward. However, if we go back to the definition of $k_{T,1}$  and $k_{T,2}$ in (\ref{rewa}) we obtain $P_{succ}(T,k_{T,1}) > P_{succ}(T,k_{T,2})$. Under assumption A we have, $k_{T,2}>k_{T,1}$. Hence the proof of  $(ii)$.\endpf\\

The last result specializes the minority game to a minority game with several thresholds allowing to control the average number of active users in each class at equilibrium. Furthermore,  we characterize how sources may design a reward mechanism in order to achieve the fairness of energy consumption based on the type of devices. For instance, if for the sake of fairness the objective is to incite more devices with high battery capacity (i.e., class 2)  to participate in forwarding compared to small devices (i.e., class 1), our scheme under mixed equilibrium may achieve this goal by assigning a reward mechanism  satisfying the relation $r_2 > \frac{r_1g_2 }{g_1}$.  In section \ref{sec:algo}  we will design a  learning algorithm that allows sources and relays to achieve the desired performance by taking into account energy consumption; the learning algorithm converges to the full mixed equilibrium without requiring perfect state information at relay nodes.

Due to the complexity of the expressions, it's in general difficult to obtain an explicit solution of (\ref{eq:mixed_hete_eqi}). We are able however to obtain numerical solution as shown in Fig.~\ref{figa}.

\begin{figure}
\center
\includegraphics[height=3in,width=0.55\linewidth,clip=]{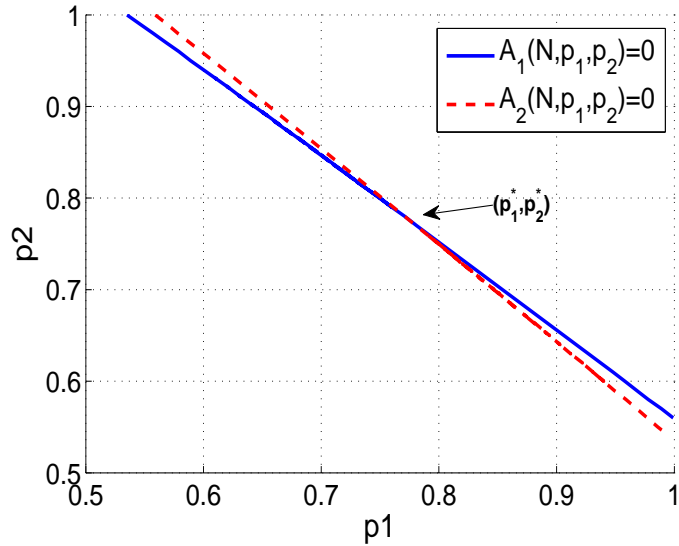}
    \caption{The mixed Nash equilibrium: multi-class, where $g1=0.8\times10^{-4},g2=0.5\times10^{-4}, r2=0.15, \lambda=0.03, \tau=100, N_1=20, N_2=20$}
    \label{figa}
    \vspace{-0.5cm}
\end{figure}

\section{Heterogeneous User-Dependent Energy Cost}\label{sec:user-dependent}

Energy depletion at DTN nodes depends not only on the wireless technology used by relays and on  the device type but also on the device's user behavior. Indeed, two users of identical devices may have drastically different energy-consumption rates: some "active" users will be draining their batteries much faster than other users. This suggests that, beyond physical characteristics of DTN nodes, the energy cost of DTN nodes should depend on the profile of users. 

In this section we focus on how the incentive mechanism combined with the minority game framework can drive this user-dependent energy cost system to an operating point that satisfies the performance requirements in the incomplete information scenario. By incomplete information we mean that, since the energy cost of a relay node is based on its activities and its behavior, sources  cannot physically identify the energy profile of a  relay node. Hence  we assume that each node  only  knows its own specific energy cost but not those of other nodes.

On the other hand sources and relay nodes have a common information on the cumulative probability distribution function $F(\cdot)$ of the energy costs in the system.  However under incomplete information, a basic approach for sources is to consider a homogeneous reward mechanism in order to achieve their target. Homogeneity is obtained by, instead of considering the cdf $F(\cdot)$, assuming that the sources will rely only on the mean $\mu$ of relays energy costs distribution.

Thus, under incomplete information, the utility function for  active node $i$ becomes
$$U_i(T)=\mathbb{E}_k \left[rP_{succ}(T,k+1) - g_i\tau \right],$$
where the expectation is taken over $k$ active relay nodes  according to binomial distribution $B(N-1, F(g))$\footnote{Recall that cdf $F(g) = P(g_i\leq g)$.}.

%

\begin{proposition}\label{prop:uedependentthresh}
 For any reward mechanism $r$, there exists a threshold-type Nash Equilibrium in which, node $i$ is active if and only if its energy cost $g_i$ is smaller than a threshold $g_i\leq g_{th}(r)$. Moreover the threshold $g_{th}(r)$ is the unique solution to $\Theta(g)=0$, where


\begin{eqnarray}
\label{zeroeq}
\Theta(g):=\underset{k=0}{\overset{N-1}{\sum}} C_{k}^{N-1} F^{k}(g) (1-F(g))^{N-k-1}\left[rP_{succ}(T,k+1) - g \tau \right]\nonumber
\end{eqnarray}
\end{proposition}
\Prf
Note that a relay node $i$  decides to be active if its utility function is positive, i.e., $U_i(T)= \Theta(g_i)\geq 0$.  Thus it is easy to check  that a relay obtains a positive utility if  its energy cost $g_i\leq g_{th}(r)$. Thus it remains to show there exists a unique solution  $g_{th}(r)$ satisfies (\ref{zeroeq}).

First we will show  the existence of solution $g_{th}$ such that  $\Theta(g)=0$.  It is easy to check that: $\lim_{g\rightarrow 0 } \Theta(g) > 0 $ and $\lim_{g\rightarrow \frac{r}{\tau} } \Theta(g) < 0$. Hence there exists a solution to $\Theta(g)=0$.
The uniqueness follows from the monotonicity of function $\Theta$. Indeed, the first term of the function $\Theta$ is decreasing function in $F(g)$ since increasing $F(g)$ engenders more active  users which decreases the probability of a relay  to receive a reward from sources (Assumption A).  Since the function $F(g)$ is increasing in  $g$, it follows that $\Theta$ is decreasing in $g$.  This ends our proof of the existence of unique solution $g_{th}$ to $\Theta(g_{th})=0$. %
\endpf

We observe that  reward  $r^*$ satisfies the following relation
\beq
 \label{eq:rewar2}
 r^{*}P_{succ}(T,\Psi)=\mu\tau,
 \eeq
 where $\Psi$ is the number of relay nodes at Nash equilibrium.  Furthermore, a relay will decide to participate if its effective energy cost is less or equal to the threshold value $g_{th}$. But the source node relies on the mean $\mu$ for the reward setting. A direct implication is that the source node may conversely provide more reward than actually needed in order to expect the targeted performance level to be achieved. This is the case when $g_{th}<\mu$. Here, expectation makes the opposite case  also possible, i.e., for a certain expected value, both larger and smaller values of the cost will be present depending on relays, which may add a fluctuation dynamic of the effectively obtained performance around the actually expected target performance. This results characterize the way how sources control the  reward mechanism under incomplete information.

\section{Distributed reinforcement learning algorithm}\label{sec:algo}


In this section we introduce a distributed reinforcement learning algorithm that permits relays to adjust their strategies over time in the framework of the DTN MG designed for heterogeneous device-dependent (section \ref{sec:netmodel}) and user-dependent (section \ref{sec:user-dependent}). Taking into account the fairness of  energy consumption, the algorithm converges to the fully mixed Nash equilibrium with no need for full state information. The analysis of convergence of the algorithm relies on a stochastic model that is associated to a continuous time deterministic dynamics. We prove that this process converges almost surely towards an  $\epsilon$-approximate Nash equilibrium.

\fdp{In DTNs, nodes' limited computational power and network's sparsity  require adaptive and distributed mechanisms letting relays adapt to operating conditions at low cost}. The learning algorithm proposed here has the following attractive features:
\begin{itemize}
\item It is genuinely distributed: strategy updating decision is local to relays;
\item It depends solely on the achieved payoffs: nodes utilize local observations to estimate their own payoffs;
\item It uses simple behavioral rule in the form of logit rule.
\end{itemize}

We assume that each relay node $i$ has a prior perception $x_i$ of the payoff performance for each action (To be active, or not), and makes a decision based on this piece of information using a random choice rule. The payoff of the chosen action is then observed and is used to update the perception for that particular action. This procedure is repeated round after round, each round of duration $\tau$ generating a discrete time stochastic process which is the learning process.

For notation's sake, denote $A=\{T,S\}$ the set of pure strategies, and $\Delta_i$ is the set of mixed strategies for player $i$ 
with $i \in \{1, . . . ,N\}$. Let $V^i(.)$ the payoff function for player $i$. The algorithm works in rounds of duration $\tau$, at round $k$, each relay node $i$ takes an action $a_i^k$  according to a mixed strategy $\pi_i^k=\sigma_i(x_i^k)\in\Delta_i$. The fully mixed strategy is generated according to the vector $x_i^k=(x_{ia}^k)_{a\in A}$ which represents its perceptions about the payoffs of the available pure strategies. In particular, relay node $i$'s fully mixed strategies are mapped from the perceptions based on the logit rule:
\beq
\label{eq:algo2}
\sigma_{ia}(x_i)=\frac{e^{\beta x_{ia}}}{e^{\beta x_{iT}}+e^{\beta x_{iS}}}
\eeq
where $\beta$ is commonly called the temperature of the logit. The temperature has a smoothing effect: when $\beta\rightarrow 0$ it leads to the uniform choice of strategies, while for $\beta\rightarrow \infty$ the probability concentrates on the pure strategy with the largest perception. We assume throughout that $\sigma_{ia}$ is strictly positive for all $a\in A$.

At round k, the perceptions $x_{ia}^k$ will determine the mixed strategies $\pi_i^k=\sigma_i(x_i^k)$ that are used by each player $i$ to choose at random action T (to be active) or S (to be silent). Then each player estimates his own payoff $\tilde{u}_i^k$, with no information about the actions or the payoffs of the other players, and uses this value ($\tilde{u}_i^k$) to update its perceptions as:
\bear
\label{eq:percep_update}
x_{ia}^{k+1}=\begin{cases}
  (1-\gamma^k)x_{ia}^k+\gamma^k \tilde{u}_i^k & \mbox{if } a_i^k=a \\
  x_{ia}^k  & \mbox{otherwise, }
\end{cases}
\eear
where $\gamma^k\in(0,1)$ is a sequence of averaging factors that satisfy $\sum_{k}\gamma^k=\infty$ and $\sum_{k}(\gamma^k)^2<\infty$ (examples of such factor are $\gamma^k =\frac{1}{k}$ or $\gamma^k =\frac{1}{1+k\log k}$). A player only changes the perception of the strategy just used in the current round and keeps other perceptions unchanged. Algorithm (\ref{alg:algorit2}) summarizes the learning process.
The discrete time stochastic process expressed in (\ref{eq:percep_update}) represents the evolution of relay node perceptions and can be written in the following equivalent form:
\beq
\label{eq:learn}
x_{ia}^{k+1}-x_{ia}^k=\gamma^k[w_{ia}^k-x_{ia}^k], \forall i\in\{1,..,N\}, a\in A
\eeq
with
\beq
w_{ia}^k=\begin{cases}
  \tilde{u}_i^k  & \mbox{if } a_i^k=a \\
  x_{ia}^k  & \mbox{otherwise.}
\end{cases}
\eeq

\begin{algorithm}[t]
\caption{Distributed reinforcement Learning Algorithm }
\label{alg:algorit2}
\begin{small}
\begin{algorithmic}[1]
\STATE\textbf{input:} $k=1$, each relay node $i$ chooses its action (T or S) according to distribution $p_i$ and set its initial perception value $x_i^0=0$.
\WHILE {$max(|x_{iT}^{k+1}-x_{iT}^k|,|x_{iS}^{k+1}-x_{iS}^k|) > \epsilon$}
\STATE Each relay node $i$ updates its fully mixed strategy profile at iteration $k$ according to (\ref{eq:algo2}).
\STATE Relay node $i$ selects its actions using its updated fully mixed strategy profile.
\STATE Relay node $i$ estimates its payoff $\tilde{u}_i^k $.
\STATE Relay node $i$ updates its perception value according to (\ref{eq:learn}).
\STATE $k \leftarrow k+1$
\ENDWHILE
\end{algorithmic}
\end{small}
\vspace{-0.08cm}
\end{algorithm}

In what follows we will prove that this algorithm can attain a steady state for the coordination process among players. Also, the information it needs to operate is minimal.
\subsection{Convergence of the Learning Process}
Based on the theory of stochastic algorithms, the asymptotic behavior of (\ref{eq:learn}) can be analyzed through the corresponding continuous dynamics \cite{Ben}:
\beq
\label{eq:ode}
\frac{dx}{dt}=E(w|x)-x,
\eeq
where $x=(x_{ia}, \forall i\in\{1,..,N\}, a\in A)$ and $w=(w_{ia}, \forall i\in\{1,..,N\}, a\in A)$.

Let us make equation (\ref{eq:ode}) more explicit by defining the mapping from  the perceptions $x$ to the expected payoff of user $i$ choosing action $a$ as $G_{ia}(x)= E(V^i|x, a_i=a)$.
\begin{proposition}
The continuous dynamics (\ref{eq:ode}) may be expressed as
\beq
\label{eq:dynam}
\frac{dx_{ia}}{dt}=\sigma_{ia}(G_{ia}(x)-x_{ia})
\eeq
\end{proposition}
\Prf Using the definition of the vector $w$, the expected value $E(w|x)$ can be computed by conditioning on player $i$'s action as:
\bear
E(w_{ia}|x_{ia})&=&\pi_{ia} U(a,\pi_{-i})+(1-\pi_{ia})x_{ia}\nonumber\\
&=&\sigma_{ia} G_{ia}(x)+(1-\sigma_{ia})x_{ia}
\eear
which with (\ref{eq:ode}) yields (\ref{eq:dynam}).
\endpf

This can be interpreted as follows: when the difference between the expected payoff and the perception value is large, the perception value, from (\ref{eq:learn}), will be updated with a large expected value $w_{ia}^k-x_{ia}^k$ and this difference will be reduced.

In the following theorem, we prove that the learning process admits a contraction structure with a proper choice of the temperature $\beta$ .

\begin{thm}
\label{thm:contract}
Under the logit decision rule (\ref{eq:algo2}), if the temperature satisfies $\beta<\frac{1}{n_s r}$, then the mapping from the perceptions to the expected payoffs $G(x)=[G_{ia}(x),\forall i\in\{1,..,N\}, a\in A)]$ is a maximum-norm contraction.
\end{thm}
\Prf
Recall that $G_{ia}(x)$ is the expected payoff of relay node $i$ choosing action $a$ given the perceptions for all players $x$. Assume the chosen action is to be active ($T$), then 
$G_{iT}(x)$ can be written as:
\[
G_{iT}(x)=\sum_{j=0}^{N} n_s rP_{succ}(T,j) C_{j}^{N}(\sigma_{iT}(x_i))^{j}(1-\sigma_{iT}(x_i))^{N-j}-g\tau
\]

Now consider the difference $G_{iT}(x_i)-G_{iT}(\hat{x}_i)$ given two arbitrary perceptions $x_i$ and $\hat{x}_i$ of a relay node $i$ :
\begin{eqnarray*}\label{eq:dif1}
|G_{iT}(x_i)-G_{iT}(\hat{x}_i)|&=&|\sigma_{iT}(x_i)\sum_{j=1}^{N-1}C_{j-1}^{N-1}(\sigma_{iT}(x_i))^{j-1}(1-\sigma_{iT}(x_i))^{N-j}U(T,j)\\
&&-\hat{\sigma}_{iT}(\hat{x}_i) \sum_{j=1}^{N-1} C_{j-1}^{N-1}(\hat{\sigma}_{iT}(\hat{x}_i))^{j-1}(1-\hat{\sigma}_{iT}(\hat{x}_i))^{N-j}U(T,j)|\\
&\leq& |\sigma_{iT}(x_i)\sum_{j=0}^{N-1}n_s r\left(C_{j}^{N-1}(\sigma_{iT}(x_i))^{j}(1-\sigma_{iT}(x_i))^{N-j}\right)\\
&&-\hat{\sigma}_{iT}(\hat{x}_i)\sum_{j=0}^{N-1}n_s r\left(C_{j}^{N-1}(\hat{\sigma}_{iT}(\hat{x}_i))^{j}(1-\hat{\sigma}_{iT}(\hat{x}_i))^{N-j}\right)|\\
&\leq& |\sigma_{iT}(x_i)n_s r-\hat{\sigma}_{iT}(\hat{x}_i)n_s r|\\
&\leq& n_s r |\sigma_{iT}(x_i)-\hat{\sigma}_{iT}(\hat{x}_i)|\nonumber
\end{eqnarray*}
 We know that $\sigma_{ia}(x_i)$ is continuously differentiable, then by the mean value theorem, there exists $\bar{x}_{ia}=\delta (x_{ia}-\hat{x_{ia}})$ with $0<\delta<1$ such that:
\bear
\sigma_{iT}(x_i)-\hat{\sigma}_{iT}(\hat{x}_i) &=& \frac{e^{\beta x_{iT}}}{\sum_{a\in A}e^{\beta x_{ia}}}- \frac{e^{\beta \hat{x}_{iT}}}{\sum_{a\in A}e^{\beta \hat{x}_{ia}}}\nonumber \\
&=& \beta\big[\frac{e^{\beta \bar{x}_{iT}}(\sum_{a \in A}e^{\beta \bar{x}_{ia}})-e^{2\beta \bar{x}_{iT}}}{(\sum_{a\in A}e^{\beta x_{ia}})^2}(x_{iT}-\hat{x}_{iT})-\sum_{a^{'}\in A, a^{'}\neq T}\beta\frac{e^{\beta \bar{x}_{ia^{'}}}e^{\beta \bar{x}_{iT}}}{(\sum_{a\in A}e^{\beta x_{ia}})^2}(x_{ia^{'}}-\hat{x}_{ia^{'}}) \big]\nonumber\\
&=&\beta\big[C_T(x_{iT}-\hat{x}_{iT})-\sum_{a^{'}\in A, a^{'}\neq T}\beta C_{a^{'}}(x_{ia^{'}}-\hat{x}_{ia^{'}}) \big] \nonumber
\eear
where $C_T=\frac{e^{\beta \bar{x}_{iT}}(\sum_{a \in A}e^{\beta \bar{x}_{ia}})-e^{2\beta \bar{x}_{iT}}}{(\sum_{a\in A}e^{\beta x_{ia}})^2}$ and $C_{a^{'}}=\frac{e^{\beta \bar{x}_{ia^{'}}}e^{\beta \bar{x}_{iT}}}{(\sum_{a\in A}e^{\beta x_{ia}})^2}$. We can easily observe $C_T=\sum_{a^{'}\in A, a^{'}\neq a} C_{a^{'}}$ and $2C_a\leq 1$. Then:
\bear
\label{eq:dif2}
|\sigma_{iT}(x_i)-\hat{\sigma}_{iT}(\hat{x}_i)|&\leq&\beta C_T |x_{iT}-\hat{x}_{iT}|+\sum_{a^{'}\in A, a^{'}\neq T}\beta C_{a^{'}}|x_{ia^{'}}-\hat{x}_{ia^{'}}|\nonumber\\
&\leq&\beta(C_T+\sum_{a^{'}\in A, a^{'}\neq T}C_{a^{'}})||x_{i}-\hat{x}_{i}||_{\infty}\nonumber\\
&\leq&  \beta||x-\hat{x}||_{\infty}.
\eear
Combining (\ref{eq:dif1}) and (\ref{eq:dif2}), we obtain
\[
|G_{iT}(x)-G_{iT}(\hat{x})|\leq \beta n_s r||x-\hat{x}||_{\infty}
\]
We obtain the same result when player $i$ chooses to be silent ($S$). Observing that since by the minority game rule $G_{iT}(\cdot)G_{iS}(\cdot) \leq 0$,
then if $\beta<\frac{1}{n_s r}$, indeed $G(x)$ is a maximum-norm contraction.
\endpf\\

Based on the property of contraction mapping, there exists a fixed point $x^*$ such that $G(x^*)=x^*$. In the following theorem we show that the distributed learning algorithm also converges to the same limit point $x^*$.

\begin{thm}\label{thm:converge}
If $G(x)$ is a $||.||_{\infty}$-contraction, its unique fixed point $x^{*}$ is a global attractor for the adaptive dynamics (\ref{eq:dynam}), and the learning process (\ref{eq:learn}) converges almost surely towards $x^*$. Moreover the limit point $x^*$ is globally asymptotically stable.
\end{thm}
\Prf Since $G(x)$ is a $||.||_{\infty}$-contraction, it admits a unique fixed point $x^*$. According to general results on stochastic algorithms the rest points of the continuous dynamic (\ref{eq:dynam}) are natural candidates to be limit point for the stochastic process (\ref{eq:learn}). All together with (\cite{Ben}, corollary 6.6), we have the almost sure convergence of (\ref{eq:learn}), given that we exhibit a strict Lyaponuv function $\phi$.\\
Now let $\phi(x)=||x_{ia}-x^{*}||_{\infty}$, then $\phi(x^{*})=0, \phi(x)>0, \forall x\neq x^{*}$. Let $i \in\{1, . . . ,N\}, a\in A$ be such that $\phi(x)=|x_{ia}-x_{ia}^{*}|$.
If $x_{ia}\geq x_{ia}^{*}$, then $\phi(x)=x_{ia}-x_{ia}^{*}$. Since $G_{ia}(x)$ is a maximum norm contraction, there exist a Lipschitz constant $\xi$ such that $G_{ia}(x)-G_{ia}(x^{*})\leq \xi(x_{ia}-x_{ia}^*)$, and $G_{ia}(x^*)=x_{ia}^*$. All together combined with equation (\ref{eq:dynam}), we can write:\\

\bear
\frac{d\phi(x)}{dt}&=&\frac{d(x_{ia}-x_{ia}^{*})}{dt}=\frac{d x_{ia}}{dt}\nonumber\\
&=&\sigma_{ia}(G_{ia}(x)-x_{ia})=\sigma_{ia}(G_{ia}(x)-G_{ia}(x^*)+x_{ia}^{*}-x_{ia})\nonumber\\
&\leq& \sigma_{ia} \xi(x_{ia}-x_{ia}^{*})+x_{ia}^*-x_{ia}=-(1-\sigma_{ia} \xi)\phi(x)<0,\forall x\neq x^*.\nonumber
\eear
and a similar argument for the case $x_{ia}\leq x_{ia}^{*}$ also shows that $\frac{d\phi(x)}{dt}<0, \forall x\neq x^*$. Thus the function $\phi(x)$ is a strict Lyaponuv function and $x^*$ is globally asymptotically stable, hence the proof.
\endpf

\subsection{Approximate Nash Equilibrium}
From lemma (\ref{thm:contract}) and theorem (\ref{thm:converge}), we have:
\[
G_{ia}(x^*)= E(V^i|x^*, a_i=a)=x_{ia}^*.
\]
This is a property of the equilibrium ($x^*$) of the distributed learning algorithm: its value $x_{ia}^*$ is an accurate estimation of the expected payoff in the equilibrium. Moreover we show that the fully mixed strategy
\[
p^*=(\sigma_{ia}^*=\frac{e^{\beta x_{ia}^*}}{e^{\beta x_{iT}^*}+e^{\beta x_{iS}^*}}, \forall a\in A, i \in \{1...N\})
\]
is an approximate Nash equilibrium.\\

\begin{proposition}
\label{prop:approx_nash}
Under the Logit decision rule (\ref{eq:algo2}), the fully mixed strategy $p^*=\sigma^*(x^*)$ at the equilibrium $x^*$ is a $\epsilon$-approximate Nash equilibrium for our game with $$\epsilon=-\frac{1}{\beta}\sum_{a\in A}\sigma_{ia}^*(ln(\sigma_{ia}^*)-1)$$ .
\end{proposition}
\Prf
A well-known characterization of the logit probabilities gives:
\bear
\sigma_{ia}(x^*)&=&\arg\max_{\sigma_i=[\sigma_{iT}, \sigma_{iS}]}\sum_{a\in A}\sigma_{ia} E(V^i|x^*,a_i=a)-\frac{1}{\beta}\sum_{a\in A}\sigma_{ia}(ln(\sigma_{ia})-1)\nonumber\\
&=&\frac{e^{\beta E(V^i|x^*, a_i=a)}}{e^{\beta E(V^i|x^*, a_i=T)}+e^{\beta E(V^i|x^*, a_i=S)}}=\frac{e^{\beta x_{ia}^*}}{e^{\beta x_{iT}^*}+e^{\beta x_{iS}^*}},\nonumber
\eear
and since (\cite{convex_opt}, pp.93)
$$
\max_{\sigma_i}\sum_{a\in A}\sigma_{ia} E(V^i|x^*, a_i=a)-\frac{1}{\beta}\sum_{a\in A}\sigma_{ia}(ln(\sigma_{ia})-1)\leq \max_{\sigma_i}\sum_{a\in A}\sigma_{ia} E(V^i|x^*, a_i=a)
$$
 then, we have:
 \[
 \sum_{a\in A}\sigma_{ia}^* E(V^i|x^*, a_i=a)\geq \max_{\sigma_i}\sum_{a\in A}\sigma_{ia} E(V^i|x^*, a_i=a)- \epsilon
 \]
 where $\epsilon=\max_{i\in \{1...N\}}\{-\frac{1}{\beta}\sum_{a\in A}\sigma_{ia}(ln(\sigma_{ia})-1)\}.$

 Hence the fully mixed strategy $p^*=\sigma^*(x^*)$ in the equilibrium $x^*$ is a $\epsilon$-approximate Nash equilibrium.
\endpf

Observe that the parameter $\epsilon$ illustrates the effect of the temperature $\beta$. A larger $\epsilon$ (smaller $\beta$) means worse learning performance.


\section{Application and Numerical Results}\label{sec:numerical}
In this section, we provide a numerical analysis of the performance achieved by DTN nodes following the distributed reinforcement learning mechanism proposed in section \ref{sec:algo}.

For the rest of the paper, we will assume that relay nodes use the two hop routing scheme, and the inter-meeting rate between nodes follows an exponential distribution. Furthermore, we assume that upon successful delivery of a message, the relay node receives a positive reward $R$ if and only if it is the first one to deliver the message to the corresponding destination.
Under these assumptions, we can obtain the expressions of different quantities: in particular the probability that an active node relays a copy of a received packet to destination within time $\tau$ is $1 - Q_{\tau}$ where the expression of $Q_{\tau}$ is given by :
$Q_{\tau} = (1 + \lambda \tau)e^{-\lambda \tau}.$
Now, the probability of successful delivery of the message for an active node is:
\beq\label{eq:condition}
\vspace{-2mm}
\!\!\!\!P_{succ}(T,k_T)=\frac{1-Q_{\tau}^{k_T}}{k_T}
\eeq
 such that each node seeks to be the first to deliver a given message to its destination(see \cite{techreport}).

The performance of our learning algorithm in the homogeneous case is shown in Fig. \ref{fig:first_2}. In this case we consider $ g=6.6\times10^{-4}, N=40$. We set the sequence $\gamma^k=\frac{1}{k}$ for all iterations $k$, and the temperature $\beta\rightarrow\infty$,  note that this choice of $\beta$ is a good deal since it allows our algorithm to attain the Nash equilibrium.

In Fig. \ref{fig:first_2} we observe that the probability to be active for a node $i$ ($p_i, \forall i \in \{1...N\}$) converges to the symmetric equilibrium ($p^*=0.35$). Moreover, it is interesting to notice that the average number of active nodes at the equilibrium approaches the value of ($\Psi=15$) where $\Psi$ defines the comfort level of the minority game in pure strategy (Fig. \ref{fig:second_2}). Such behavior is, in fact, a convergence to the strictly fully mixed Nash equilibrium discussed in proposition (\ref{sym}).
\begin{figure}
    \centering
    \caption{Learning the fully mixed strategy: homogeneous case. $g=6.6\times10^{-4}$}
        \includegraphics[height=1.9in,width=0.55\linewidth,clip=]{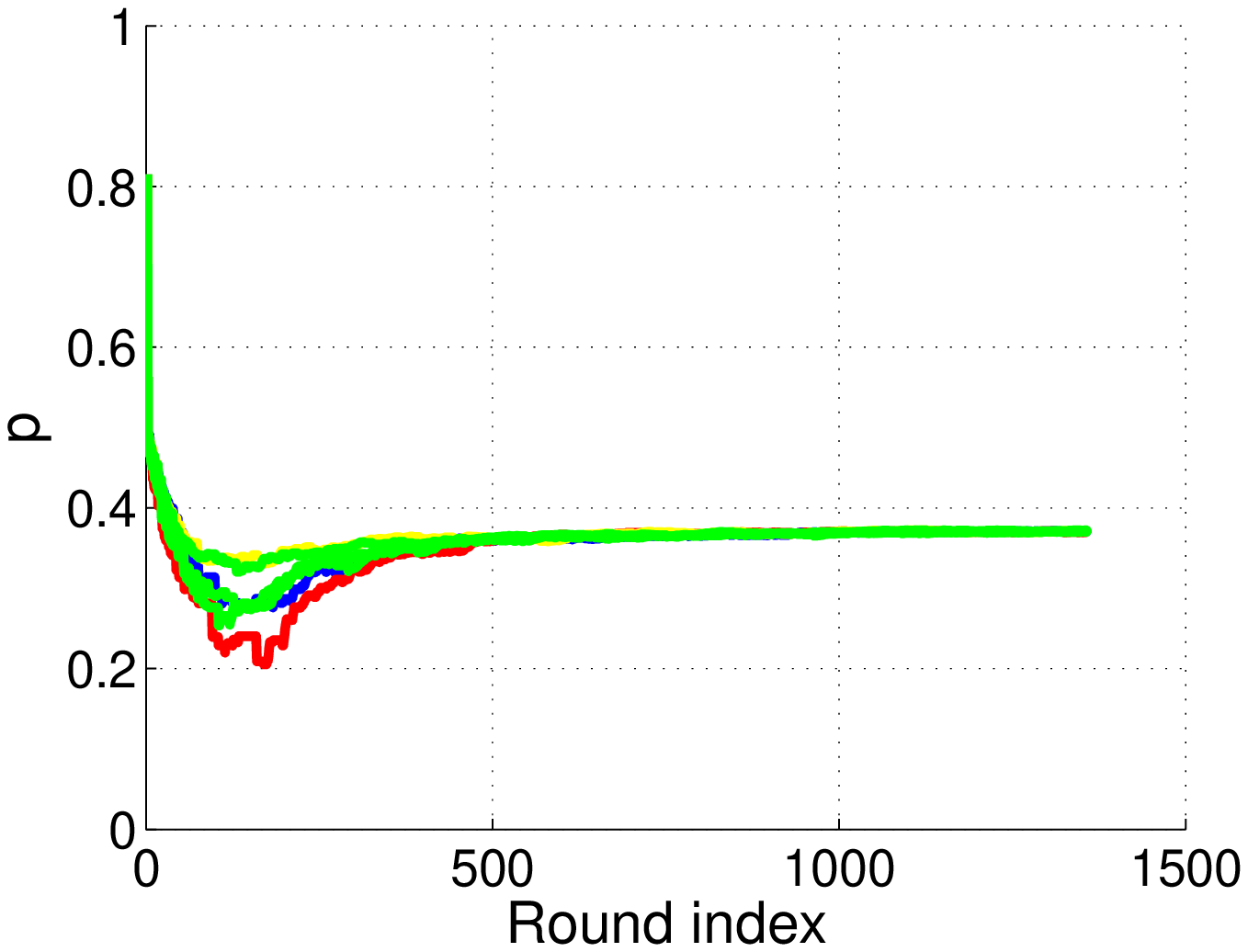}
        \label{fig:first_2}
\end{figure}

\begin{figure}
\centering
        \caption{Learning the fully mixed strategy: homogeneous case. $g=6.6\times10^{-4}$}
        \includegraphics[height=1.9in,width=0.55\linewidth,clip=]{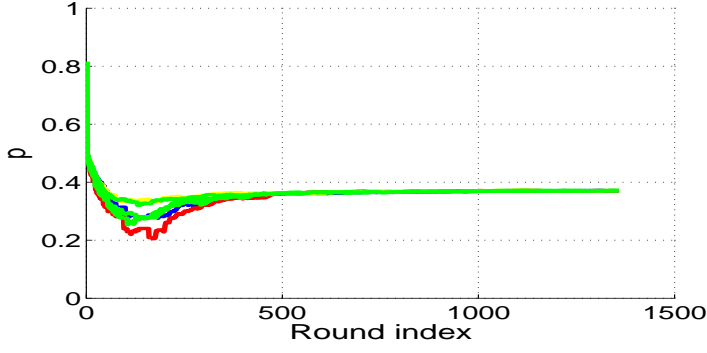}
        \label{fig:second_2}
\end{figure}

Now we examine the performance of our algorithm in a multi-class framework (device-dependent heterogeneity), where we consider the existence of two classes of nodes.
The parameters $\lambda=0.03, \tau=100$ are used through out the numerical analysis.

The performance of the learning algorithm in the heterogeneous DTN is investigated in two cases, symmetric (i.e. when $\frac{g_1}{r_1} = \frac{g_2}{r_2}$) and asymmetric ($\frac{g_1}{r_1} \neq \frac{g_2}{r_2}$). We consider first the symmetric case. We consider first the symmetric case . We consider $g_1=0.8\times10^{-4},g_2=0.5\times10^{-4}, N_1=20, N_2=20$ then setting $r2=0.15$ we obtain $r_1=0.24$.
In Fig. (\ref{fig5})(a) we observe that the probability of being active of nodes of both classes ($p_1,p_2$) converges to the symmetric Nash equilibrium discussed in proposition (\ref{mixed_equi_multi_case}), and the value it converges to ($p_1^*=p_2^*=0.78$) is the solution of the equation ($A_1(N,p_1^*,p_2^*)=A_2(N,p_1^*,p_2^*)=0$). The average number of active nodes, depicted in Fig (\ref{fig5})(b), converges to $\Psi=30$ that satisfies the relation (\ref{eq:equilibre}).

\begin{figure}
    \centering
    \subfigure[ ]
    {
        \includegraphics[height=1.9in,width=0.48\linewidth,clip=]{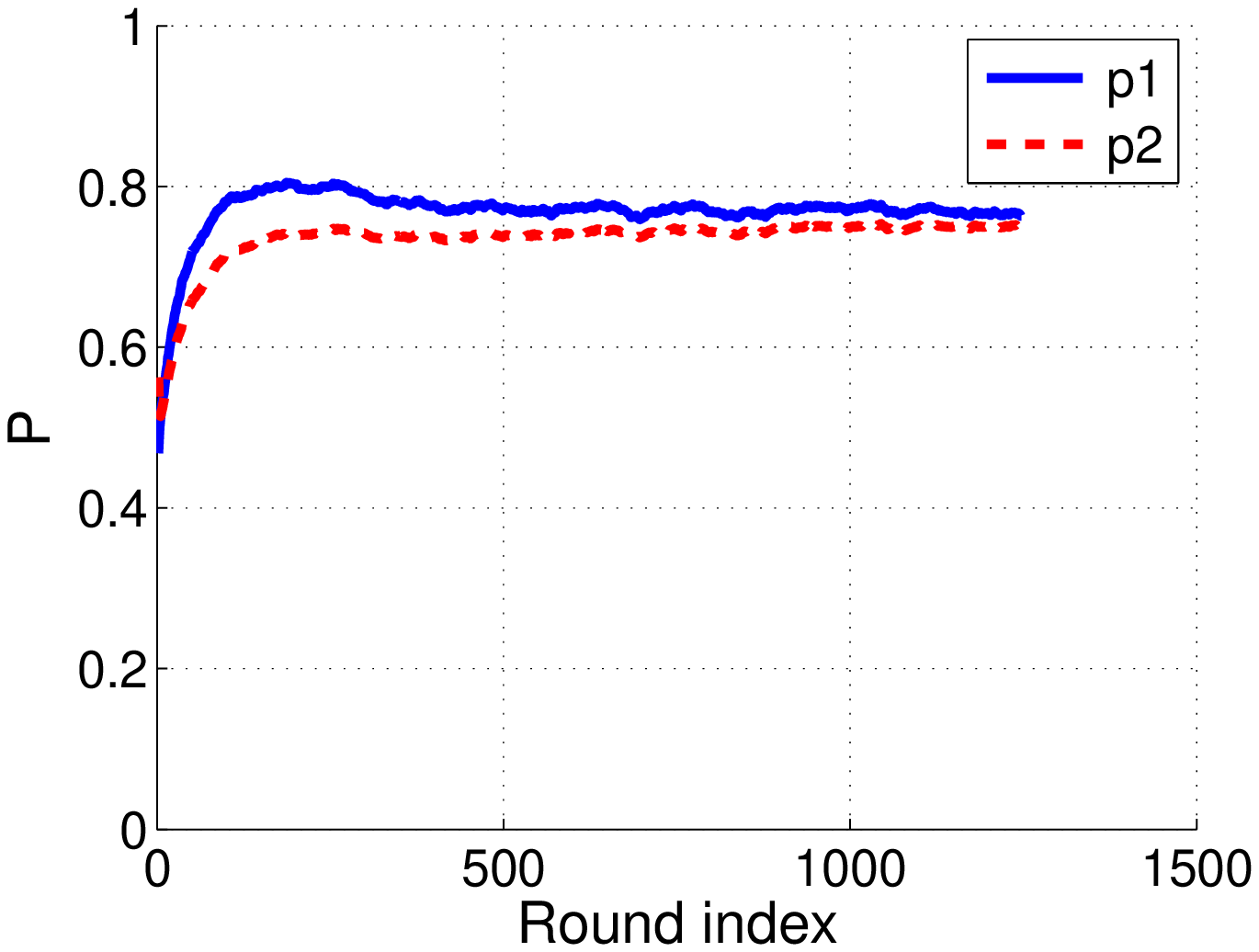}
        \label{fig:first_2}
    }
        \subfigure[]
    {
        \includegraphics[height=1.9in,width=0.48\linewidth,clip=]{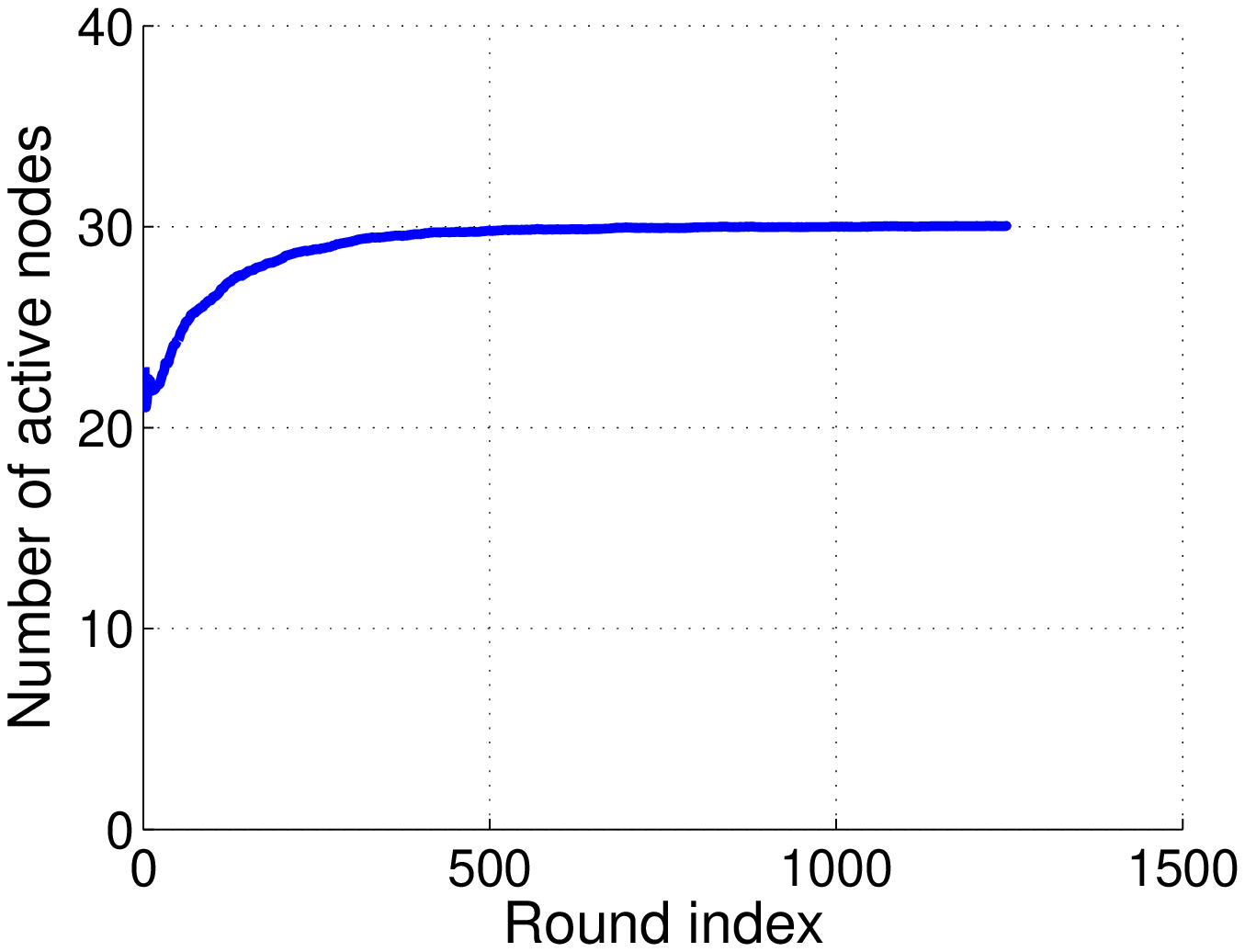}
        \label{fig:second_2}
    }

       \caption{Learning the fully mixed strategy: heterogeneous symmetric case, where: $g1=0.8\times10^{-4},g2=0.5\times10^{-4},r2=0.15$.}
    \label{fig5}

\end{figure}

In Fig(\ref{fig6}), we depict the asymmetric case, when $g_1>g_2$ and $r_1 < \frac{g_1 r_2}{g_2}$. In Fig(\ref{fig6})(a) we observe that ($p_2>p_1$), in other words, the nodes with high energy constraint (class $1$) are less active, thus by allocating smaller reward ($r_1$), fewer nodes of class $1$ are active. Notice in Fig(\ref{fig6})(b) that the average number of active nodes $\Psi_1\leq k_T<\Psi_2$.

\begin{figure}
    \centering
     \subfigure[ ]
    {
        \includegraphics[height=1.9in,width=0.48\linewidth,clip=]{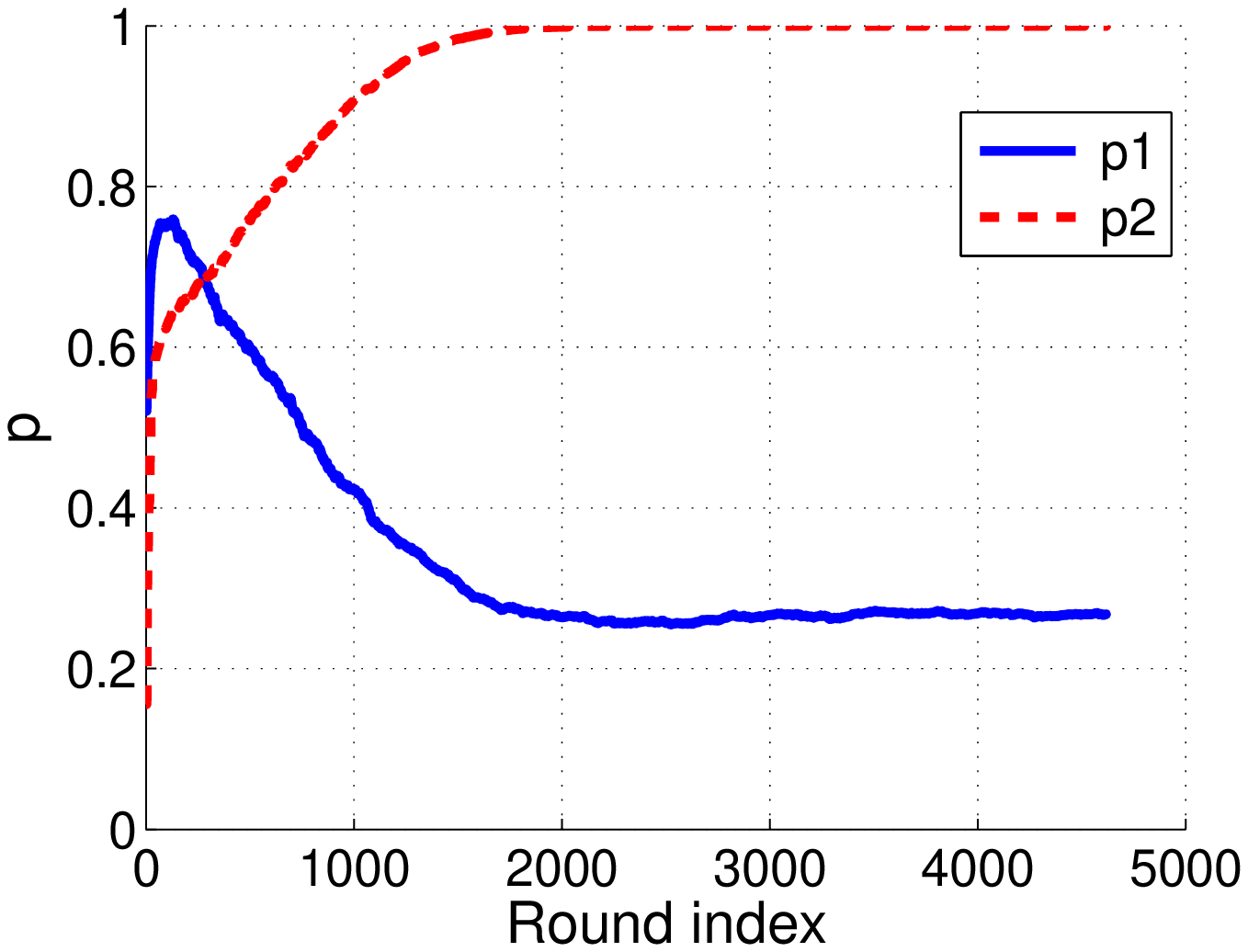}
        \label{fig:first_2}
    }
        \subfigure[]
    {
        \includegraphics[height=1.9in,width=0.48\linewidth,clip=]{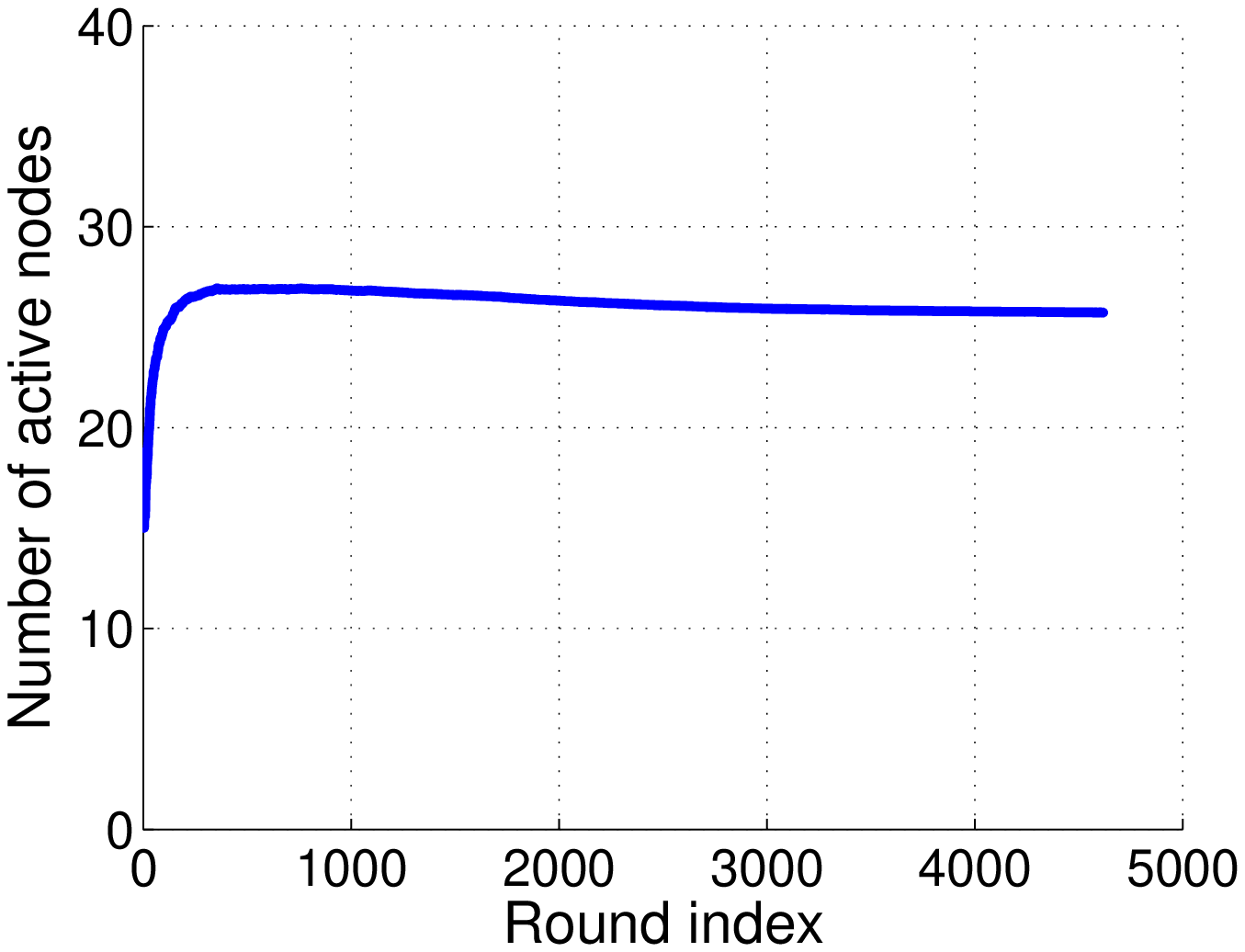}
        \label{fig:second_2}
    }
       \caption{Learning the fully mixed strategy: heterogeneous asymmetric case, where: $g1=0.8\times10^{-4},g2=0.5\times10^{-4}, r_1=0.2, r_2=0.14$ which yields $\Psi_1=26, \Psi_2=28. $}
    \label{fig6}
  \end{figure}

\begin{figure}
    \centering
     \subfigure[ ]
    {
        \includegraphics[height=2.1in,width=0.45\linewidth,clip=]{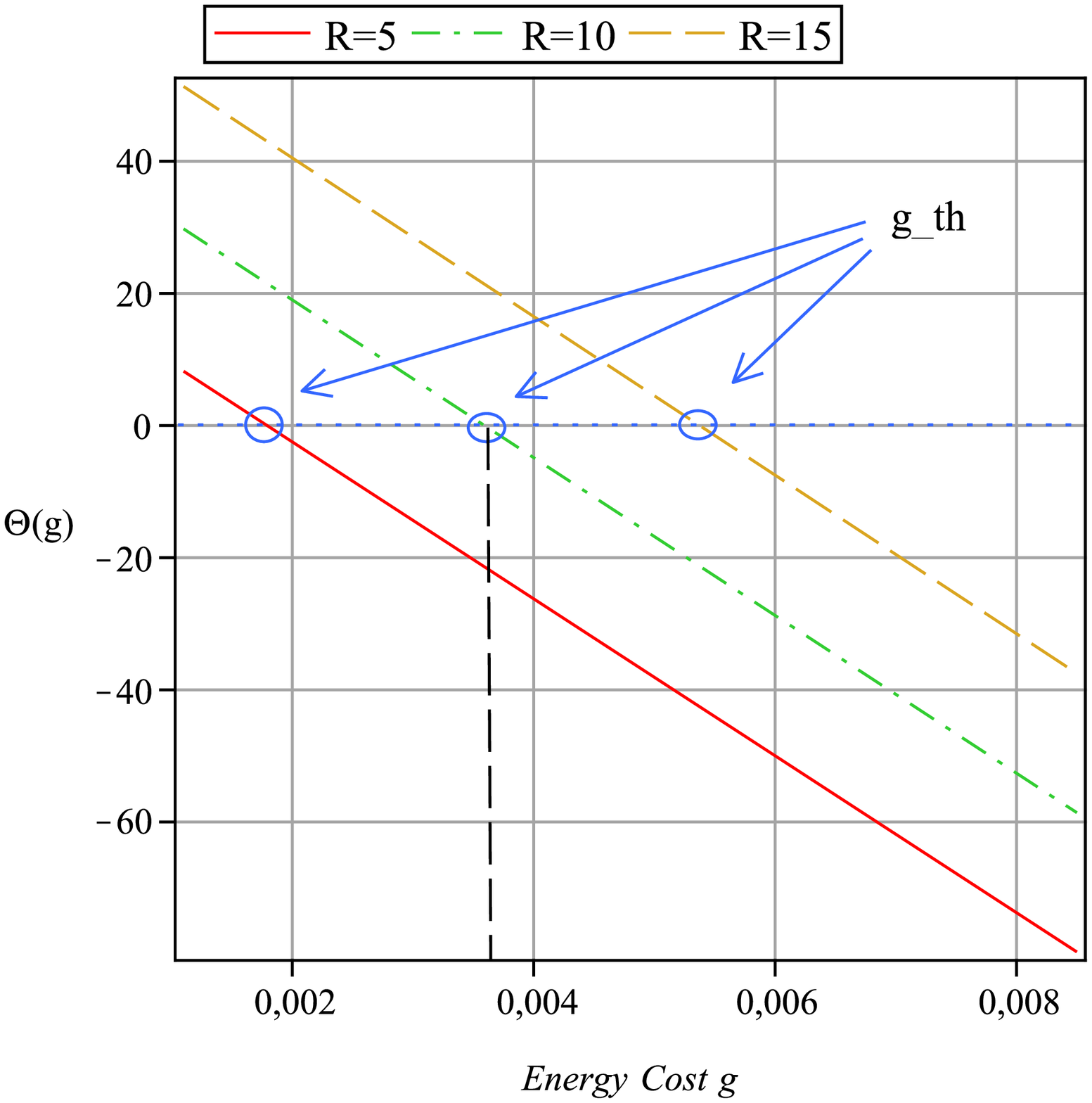}
        \label{fig:first_2}
    }
        \subfigure[]
    {
        \includegraphics[height=2.1in,width=0.45\linewidth,clip=]{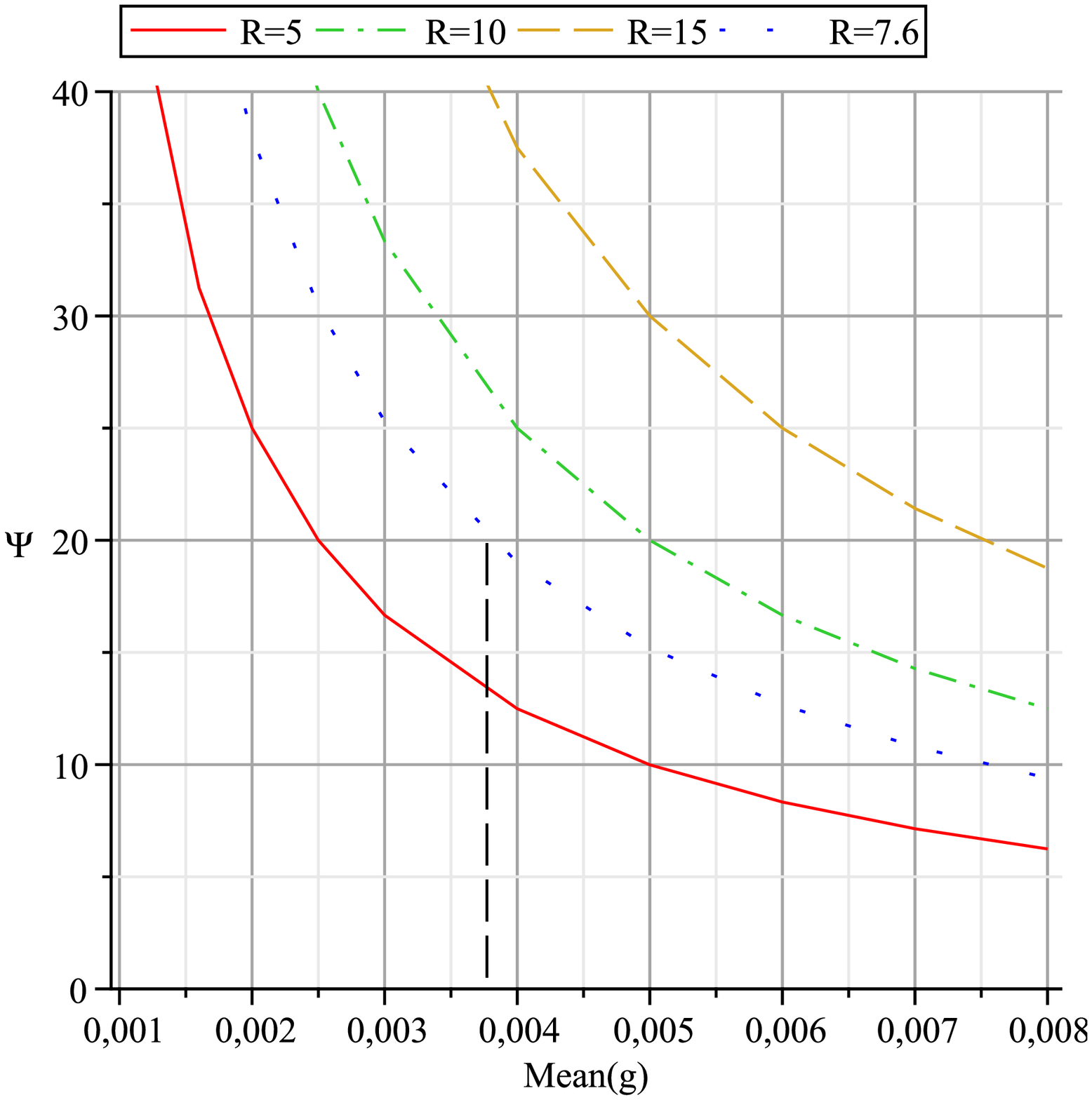}
        \label{fig:second_2}
    }
       \caption{On the left, $\Theta(g)$ as a function of $g$ and $R$. On the right, $\Psi$ as a function of $\mu$ and $R$ , where: $N=40, \tau=100, \lambda=0.03$.}
    \label{fig7}
  \end{figure}

A summarized numerical validation of our results for the user-dependent incomplete-information scenario is depicted in Fig.(\ref{fig7}). In Fig.(\ref{fig7},a), we show existence of the equilibrium threshold (\ref{prop:uedependentthresh}) and several of its characteristics. First, we note that function $\Theta(g)$ (\ref{zeroeq}) is decreasing in the users\rq{} energy cost $g$ and increasing with reward $R$. Second, the equilibrium threshold $g_{th}$ is also increasing in $R$, which means that the individual incitement turns into a global incitement despite competition. Third, in Fig.(\ref{fig7},b), we show the reward setting from the source's perspective, putting into relation the targeted number of active relays with expected mean value on users energy costs. Through a joint analysis of Fig.(\ref{fig7},a) and Fig.(\ref{fig7},b) it is interesting to notice that since the source only knows the mean $\mu$, it might pay more reward to achieve a performance level that could have been reached with less reward. Indeed, let\rq{}s assume for example, the case $\Psi=20$. Given $\mu=0.005$ and according to Fig.(\ref{fig7},b) the source needs to set a reward $R=10$ to incite $20$ relay nodes to be active. However, relying on  Fig.(\ref{fig7},a), for this case we have $g_{th}=0.0038$, hence the source could have reached the same target number $\Psi=20$ with a reward $R=7.6$.
%

\section{Discussion of Assumptions, Limitations and Future Work}
\label{sec:discuss}

In this section we discuss the main assumptions that were adopted to yield a tractable model and we describe limitations and possible extensions.

{\noindent \bf Mobility pattern}:  A key challenge in developing our results has been to make general assumptions about the mobility of DTN nodes. In particular,  the properties derived for our incentive mechanism hold under any homogeneous mobility pattern. Indeed, the large majority of analytical studies are based on some assumptions on the mobility  for the sake of tractability.  Early works typically assumed that the  cumulative distribution function of  inter contact time  decays exponentially over time such as in random waypoint models.  However, extensive empirical mobility traces later show that the CCDF of the inter-contact time follows approximately a power law over large time range with exponent less than unit \cite{chaintreau07impact}.

A further assumption to prove convergence of our stochastic approximation algorithm is that nodes are identical and uniformly visit the entire network space. Experimental data, however, have shown that mobility patterns of individuals are typically restricted to a given area, and the overall node density is often largely inhomogeneous.  Such models allow studying how DTN routing mechanisms are affected by highly inhomogeneous node density, diverse mobility patterns and transmission technologies. In future work, we will adapt our mechanism for cases where heterogeneity is caused under human mobility during a day.  We will study how to model such changes and how routing algorithms can take into account these time-of-the-day effects.  A possible initial model would account for users moving in a periodic manner over a daily interval, thereby creating some heterogeneity in both time and space.

{\noindent  \bf Buffer management}:
In our model, we assume that relay nodes have enough capacity to store messages generated from sources and their copies.  But it  is clear that, in the context of DTNs, node buffers may well overflow if no message discarding policy is adopted. In turn, performance measures for DTNs depend not only on the number of active relay nodes but also on buffer capacity.  In this scenario,  efficient drop policies at relay nodes  decide which messages should prioritized under capacity constraints regardless of the specific routing algorithm used.

{\noindent \bf Delivery probability}:
A central  performance measure  studied in DTNs literature is delivery probability. This measure  holds relevant  for several applications which may be running over DTNs. However our scheme can be designed to attain general performance metrics provided that they satisfy  assumption {\bf A}; for instance, end-to-end delay is such a metric.

{\noindent \bf Routing and Protocols}: We did not address specific protocols for the delivery of messages, rather we have considered an incentive scheme where only successful relays receive a reward. In turn, this scheme is indeed general and the proposed framework in principle can work for any DTN routing protocol.  But, the assumption about how a relay node obtains a reward may limit coordination between relay nodes.  In fact,  in  order to avoid the use of feedbacks that allow relays to know  whether the message has been successfully delivered or not, we assume that a relay will receive a reward if and only if it is the first one to deliver the message.  Unfortunately, this reward scheme may foster unintended deviant behaviors in which relay nodes may refuse to forward messages to other relays in order to increase their own utility \cite{Bitcoin}. In the future, we will propose a modification of our mechanism that can eliminate this problem by using some additional rewards  between relays in order to incentivize messages propagation  between them \cite{Bitcoin}.



\section{Conclusion}\label{sec:concl}


In DTNs composed by mobile nodes, not always users will be willing to forward a message to destination. In fact, it is reasonable
 that they disregard the service as DTNs relays in order to preserve batteries. To this respect, the design of an efficient activation control
becomes a vital requirement for any such communication system.

 In this paper we have devised a rewarding scheme where relays gain certain rewards
that incentive them to sacrifice  memory and battery on DTNs relaying operations. Furthermore, we make a specific effort
such in a way that our mechanism is designed to account for heterogeneous user profiles and devices energy costs.

Furthermore we argue that  any such a coordination scheme should not rely on end to end control message exchange.
DTNs characteristics in fact discourage the usage of timely feedback to enforce cooperative schemes which may
 be implemented on mobile nodes. This is a seemingly fundamental obstacle in order to permit coordination of
 such systems. To this respect, our paper provides a novel key contribution: the reward mechanism in fact is
 designed using the theory of  Minority Games (MGs) in order to attain coordination in distributed fashion.  Overall,
 our scheme covers several possible  information scenarios that sources and relay nodes may face in reality, ranging from
full state information to imperfect state information and applies to general intermeeting distributions for nodes contacts.

Also, in order to prove the correctness of our incentive mechanism, we have provided
a complete  characterization of the equilibria of the baseline MG in the case of heterogeneous DTNs.
Finally, the core machinery to attain feedback-less coordination is based on a learning algorithm that involves
stochastic approximations: our algorithm provably drives the system to the aforementioned equilibria while
requiring just local estimations of system parameters performed at mobile relay nodes.

 \bibliographystyle{abbrv}
\bibliography{mybib}  

\end{document}